\documentclass[usenatbib]{mn2e}
\usepackage{fixltx2e,longtable,amsmath,amssymb}
\usepackage[pdftex]{graphicx,color}

\newcommand{\Teff}{\ensuremath{T_{\rm eff}}}
\newcommand{\logg}{\ensuremath{\log{g}}}
\newcommand{\mdot}{\ensuremath{\dot{M}}}
\newcommand{\msun}{\ensuremath{\mbox{M}_{\odot}}}
\newcommand{\rsun}{\ensuremath{\mbox{R}_{\odot}}}
\newcommand{\lsun}{\ensuremath{\mbox{L}_{\odot}}}

\newcommand{\rstar}{\ensuremath{R_*}}
\newcommand{\vinf}{\ensuremath{v_\infty}}

\newcommand{\vesini}{\ensuremath{v_{\rm e}\sin{i}}}

\newcommand{\Ha} {\ensuremath{\mbox{H}\alpha}}
\newcommand{\Hb} {\ensuremath{\mbox{H}\beta}}

\newcommand{\Wl} {\ensuremath{W_\lambda}}

\newcommand{\hei} {\mbox{He$\;$\sc{i}}}
\newcommand{\heii} {\mbox{He$\;$\sc{ii}}}
\newcommand{\oii} {\mbox{O$\;$\sc{ii}}}
\newcommand{\siiii} {\mbox{Si$\;$\sc{iii}}}
\newcommand{\siiv} {\mbox{Si$\;$\sc{iv}}}
\newcommand{\siv} {\mbox{S$\;$\sc{iv}}}
\newcommand{\cii} {\mbox{C$\;$\sc{ii}}}
\newcommand{\ciii} {\mbox{C$\;$\sc{iii}}}
\newcommand{\civ} {\mbox{C$\;$\sc{iv}}}
\newcommand{\nii} {\mbox{N$\;$\sc{ii}}}
\newcommand{\niii} {\mbox{N$\;$\sc{iii}}}
\newcommand{\nv}  {\mbox{N$\;$\sc{v}}}

\newcommand\5{{\footnotesize V}}

\newcommand\3{{\footnotesize III}}
\newcommand\2{{\footnotesize II}}

\newcommand{\kms} {\ensuremath{\mbox{km}\;\mbox{s}^{-1}}}

\title[Spectroscopy of HD 191612]
{Towards an understanding of the Of?p star HD~191612:  optical spectroscopy}

\author[I.\ D.~Howarth et al.]  
{Ian D.\ Howarth,$^1$\thanks{email: idh@star.ucl.ac.uk}
Nolan R.~Walborn,$^2$
Danny J.~Lennon,$^3$
Joachim~Puls,$^4$
 Ya\"el~Naz\'e,$^5$\newauthor 
K.~Annuk,$^6$ 
I.~Antokhin,$^7$ 
D.~Bohlender,$^8$ 
H.~Bond,$^2$ 
J.-F.~Donati,$^9$ 
L.~Georgiev,$^{10}$\newauthor
D.~Gies,$^{11}$ 
D.~Harmer,$^{12}$ 
A.~Herrero,$^{13}$ 
I.~Kolka,$^6$  
D.~McDavid,$^{14}$
T.~Morel,$^{15}$ \newauthor
I.~Negueruela,$^{16}$ 
G.~Rauw,$^5$  
P.~Reig$^{17}$\\
$^1$Dept.\ of Physics and Astronomy, University College London,  Gower Street, London WC1E 6BT, UK  \\ 
$^2$Space Telescope Science Institute, 3700 San Martin Drive, Baltimore, MD~21218, USA\\ 
$^3$The Isaac Newton Group of Telescopes, Apartado de Correos 321, 38700 Santa Cruz de La Palma, Canary Islands, Spain\\
$^4$Universit\"ats-Sternwarte M\"unchen, Scheinerstr. 1, 81679 M\"unchen, Germany\\
$^5$FNRS Institut d'Astrophysique et de G\'{e}ophysique, Universit\'{e} de Li\`{e}ge, B\^{a}t.~B5c, All\'{e}e du VI Ao\^{u}t~17, B-4000 Li\`{e}ge, Belgium  \\
$^6$Tartu Observatory, 61602 T\~oravere, Estonia\\
$^7$Sternberg Astronomical Institute, Moscow University, Universitetskij Prospect~13, Moscow~119992, Russia\\
$^8$National Research Council of Canada, Herzberg Institute of Astrophysics, 5071 West Saanich Road, Victoria, BC~V9E~2E7, Canada\\
$^9$Laboratoire d'Astrophysique, Observatoire Midi-Pyr\'{e}n\'{e}es, 14~Av.~E.~Belin, F-31400 Toulouse, France\\
$^{10}$Instituto de Astronom\'{\i}a, UNAM, Apartado Postal 70-254, CD Universitaria, CP 04510 M\'{e}xico DF, Mexico\\
$^{11}$Department of Physics and Astronomy, Georgia State University, P.O.~Box~4106, Atlanta, GA~30302-4106, USA\\
$^{12}$Kitt Peak National Observatory, NOAO, P.O.~Box~26732, Tucson, AZ~85719, USA\\
$^{13}$Instituto de Astrof\'{\i}sica de Canarias, 38200 La Laguna, Tenerife, Spain\\
$^{14}$Department of Astronomy, University of Virginia,    P.O.~Box~400325,   Charlottesville, VA~22904-4325, USA\\
$^{15}$Katholieke Universiteit Leuven, Departement Natuurkunde en
Sterrenkunde, Instituut voor Sterrenkunde, Celestijnenlaan 200D, B-3001 Leuven, Belgium\\
$^{16}$Departamento de Física, Ingenier\'{\i}a de Sistemas y Teor\'{\i}a de la Se\~{n}al, Escuela Polit\'{e}cnica Superior, Universidad de Alicante, Ap.~99,\\ 03080 Alicante, Spain\\
$^{17}$IESL, Foundation for Research and Technology, 71110 Heraklion, Greece\\
}

\date{Dates to be inserted}

\begin{document}

\maketitle

\begin{abstract}
We present extensive optical spectroscopy of the early-type magnetic
star HD~191612 (O6.5f?pe--O8fp).  The Balmer and \hei\ lines show
strongly variable emission which is highly reproducible on a
well-determined \mbox{538-d} period.  Metal lines and \heii\
absorptions (including many selective emission lines but excluding
\heii~$\lambda$4686{\AA} emission) are essentially constant in line
strength, but are variable in velocity, establishing a double-lined
binary orbit with $P_{\rm orb}= 1542$d, $e=0.45$.  We conduct a
model-atmosphere analysis of the primary, and find that the system is
consistent with a $\sim$O8 giant with a $\sim$B1 main-sequence
secondary.  Since the periodic \mbox{538-d} changes are unrelated to
orbital motion, rotational modulation of a magnetically constrained
plasma is strongly favoured as the most likely underlying `clock'.
An upper limit on the equatorial rotation is consistent with this
hypothesis, but is too weak to provide a strong constraint.
\end{abstract}

\section{Introduction}
Peculiarities in the spectrum of the early-type star HD~191612 were
first noted by \citet{Walborn73}; it remains one of only three known
Galactic examples of the Of?p class,\footnote{The others are HD~108
  \&\ HD~148937; the SMC Of?p stars AzV~220 \&\ 2dFS~936 were found
  subsequently \citep{Walborn00, Massey01, Evans04}.}  this
designation indicating C$\;${\sc iii}~$\lambda$4650 in emission with
comparable strength to N$\;${\sc iii}~$\lambda$4640. Renewed interest
followed the discovery of recurrent spectral variability between
spectral types O6--O7 and O8 (with correlated changes in the unusual
emission-line features; \citealt{Walborn03}), which \citet{Walborn04}
showed to be consistent with a \mbox{$\sim$540-d} period identified in
{\em Hipparcos} photometry \citep{Koen02, Naze04}.

From timescale arguments, \citet{Walborn04} suggested that the `clock'
underlying the variability was most probably a binary orbit.  New
light was cast on this issue by \citeauthor{Donati06a}
(\citeyear{Donati06a}a), who found HD~191612 to be only the second
O-type star known to possess a magnetic field.\footnote{The first was
  $\theta^1$~Ori~C; \citet{Donati02}.}  Although their observations
sampled only a single epoch, they were none the less able to estimate
a polar field strength of $\sim$1.5kG (from an observed line-of-sight
field of $\sim$220G, by assuming a dipole field), and made the case
for \mbox{538-d} {\em rotational} modulation, arguing that the field
itself could easily be responsible for the implied slow rotation
(through magnetic braking).

Of observational necessity, the magnetic, rotational-modulation model
is as yet poorly constrained, and the discussion of XMM-Newton
spectroscopy by \citet{Naze07} emphasizes a number of discrepancies
with the X-ray behaviour expected in the simplest version of this
scenario.  On the other hand, the alternative orbital model has not
been subject to any strong tests (arguably, even a strict, coherent
spectroscopic periodicity has yet to be demonstrated robustly).  Here
we present the results of an extensive campaign of optical
spectroscopy, carried out in an attempt to shed light on these issues.

\section{Observations}

The major part of our campaign was conducted during the 2004 and 2005
observing seasons.  Because of the $\sim$18-month variability
timescale, scheduled observations at common-user facilities were
generally impractical, and our observations were obtained through
service programmes; by taking advantage of telescope time awarded to
other scheduled programmes; and by exploiting the goodwill of
colleagues.  The main dataset, summarized in Table~\ref{tab_obs} (166
digital observations spanning 17 years), is therefore quite
heterogeneous.  None the less, the spectra can be conveniently
characterized by wavelength range (`red', including
H$\alpha$~6563{\AA}; `blue', generally including at least the
$\sim$4400--4700{\AA} region) and by resolution (`high', $R\ga 4\times
10^4$; `intermediate', $R\ga 4\times 10^3$; and `low').  With a few
exceptions, the spectra are generally reasonably well exposed, with
signal:noise ratios typically approaching $\sim$100.

All spectra have been put on a helio{\-}centric velocity scale, and
all the H$\alpha$ spectra have been corrected for telluric absorption
by division, in topocentric space, with an appropriately scaled and
smoothed telluric `map' constructed from high-quality, high-dispersion
echelle spectra.  (Because the telluric lines are unresolved, direct
scaling in optical depth is impossible; we scaled in observed
intensity, but checked that scaling in observed, pseudo-optical depth
gives negligibly different results.)

\section{The 538-d period:  H$\alpha$ variability}

As already noted by \citet{Naze07}, the spectral lines can, for the
most part, be separated into two groups according to variability
characteristics: the absorption lines of metals and of \heii\ show, at
most, small changes in line strength, while the hydrogen and \hei\
lines show large equivalent-width changes.  H$\alpha$ shows the
largest-amplitude variations of any spectral feature followed in our
campaign, by a comfortable margin.  This, and the rather extensive
temporal coverage of the red-region spectra, mean that it is the most
useful feature for investigating the periodicity of the spectroscopic
variability.

\subsection{Ephemeris}
\label{sec_HaP}

Even casual inspection shows systematic variations in H$\alpha$ that
repeat on a period close to the $\sim$540-d timescale found in {\em
  Hipparcos} photometry (over only about 2 cycles).  In order to
quantify that period and its uncertainty, we fit an {\em ad hoc}
analytical function to the equivalent-width measurements; we find that
a truncated gaussian,
\begin{align}
\Wl(\phi_\alpha) &=  
W_0 - A \exp\left\{{\frac{-\phi_\alpha^2}{2 \sigma^2_\phi}}\right\}
&-\phi_0 < \phi_\alpha < +\phi_0 
\nonumber 
\\
&=   
\Wl(\phi_0)
\phantom{W_0 - A \exp\left\{{\frac{-\phi_\alpha^2}{2 \sigma_\phi}}\right\}}
&
\phi_0 \le \phi_\alpha \le 1-\phi_0
\label{eqn_fform}
\end{align}
gives a good match to the observations.  Although this functional form
is arbitrary, it does provide a useful characterization of the data;
the best-fit parameters are listed in Table~\ref{tab_ha}, and lead to
the ephemeris
\begin{align}
\phi_\alpha = (t - \mbox{JD}\;2453415.2)/537.6\mbox{d}
\label{eqn_ephem}
\end{align}
where phase zero corresponds to peak \Ha\ emission.\footnote{This
  differs from the \citet{Walborn04} ephemeris (which was tied to the
  less well determined minimum in the {\em Hipparcos} photometry) by
  half a cycle.}

The phase-zero epoch is chosen to be close to the median date of the
observations (to minimize the formal error on $t_0$), and so, because
the density of observations has increased with time, many spectra were
obtained at negative epochs.  For such observations we use a
logarithmic-like notation for phases, such that $\phi_\alpha =
\overline{1}.23$ means phase +0.23 in cycle $-1$.

\begin{table}
\caption[]{\Ha\ variability: best-fit parameters for the arbitrary functional
form described by eqtn~\ref{eqn_fform}.}
\begin{tabular}{lrcll}
$W_0$ & 2.51         &$\pm$& 0.18&\AA\\
$A$ & 6.74                && 0.16&\AA\\
$P_\alpha$ & 537.6               && 0.4 &d\\
$t_0$ & JD$\;$2,453,415.2 && 0.5 &\\
$\sigma_\phi$ & 0.177     && 0.005&\\
$\phi_0$ &   0.337        && 0.005\\
\end{tabular}
\label{tab_ha}
\end{table}

\subsection{\Ha\ Properties}
\label{sec_Ha}

\begin{figure*}
\center{\includegraphics[scale=0.65,angle=-90]{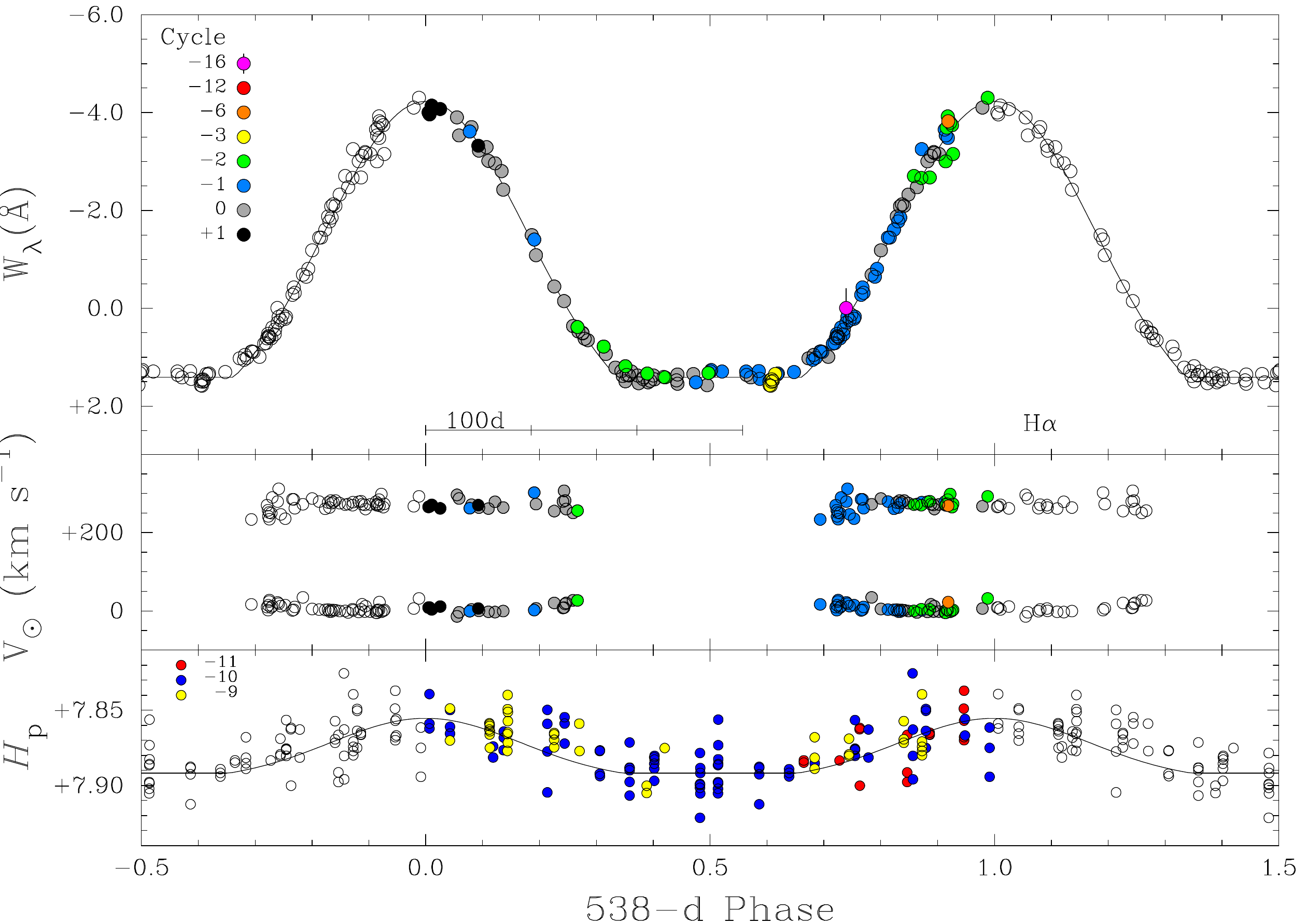}} 
\caption[]{\Ha\ measurements folded on the ephemeris of
eqtn.~\ref{eqn_ephem}, shown over two cycles.
\textit{Upper panel:}  equivalent width (the solid line is the {\em ad hoc}
functional fit, eqtn.~\ref{eqn_fform}, described in
Section~\ref{sec_HaP}).
\textit{Middle panel:} FWHM
(upper groups of points) and central velocity (lower) of excess emission.
\textit{Bottom panel:} the {\em Hipparcos} photometry (with a scaled, 
vertically shifted
version of
the \Ha\ functional fit to guide the eye).}
\label{fig_Ha}
\end{figure*}

\begin{figure}
\center{\includegraphics[scale=0.5,angle=0]{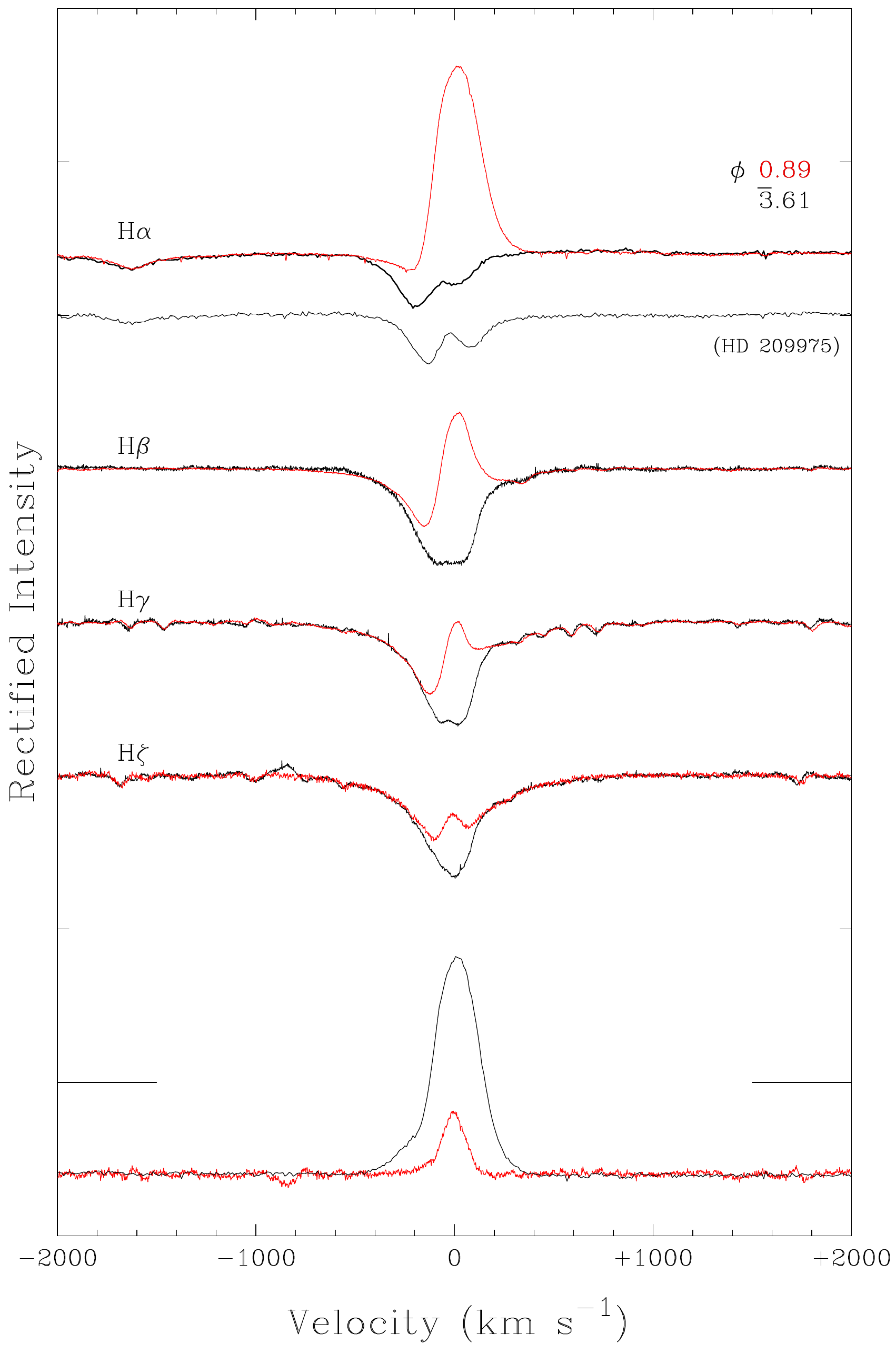}} 
\caption[]{Balmer-line profiles for HD~191612
at quiescence ($\phi=\overline{3}.61$, in black) and near emission maximum
($\phi=0.89$, red);
tickmarks on the $y$ axis are separated by half the continuum level.
The \Ha\ spectrum of HD~209975 is included to demonstrate that a near match
to the
minimum-state spectrum can be found among normal stars
(see Section~\ref{sec_Ha} for details).
Difference
spectra for \Ha\ and H$\zeta$ are shown at the bottom of the plot.
Velocities are helio{\-}centric.}
\label{fig_balmer}
\end{figure}

Figure~\ref{fig_Ha} shows the \Ha\ equivalent-width measurements,
which vary between +1.4 and $-4.3$\AA, folded on the adopted
ephemeris; it illustrates a number of noteworthy points:
\begin{enumerate}
\item The \Ha\ light-curve is remarkably symmetrical about phase zero.
\label{en1it1}
\item There is a well-defined interval of 
apparent `quiescence', lasting 
$\sim{0.3}(=1\mbox{--}2\phi_0)$ of the period.
\item The behaviour is repeatable
from the earliest quantitative data (1982),\footnote{We 
measured the equivalent width from a 
digitized version of the
plot of \Ha\ given by \citet{Peppel84}, who observed on
JD 2,445,211, 24 years before our last observation.  We included this point
in the fit of eqn.~\ref{eqn_fform}, but its exclusion makes no
important changes to the ephemeris.} supporting a
truly periodic underlying \mbox{538-d} `clock'.
\label{en1it2}
\item None the less, the \Ha\ profiles are evidently not {\em
    strictly} repeatable; the standard deviation of \Wl\ measurements
  about the functional fit is only 0.16\AA, but scrutiny of the
  spectra shows that real, small-amplitude variability contributes to
  this dispersion at all sufficiently well-sampled phases, including
  quiescence, on timescales longer than a few days.  We suspect that
  any O-type star observed as extensively and intensively as HD~191612
  may show \Ha\ `jitter' at a similar level (cf., e.g.,
  \citealt{Kaper97, Morel04}).
\label{en1it3}
\end{enumerate} 
The \Ha\ profile during quiescence is asymmetric and filled in by
stellar-wind emission (Fig.~\ref{fig_balmer}).  It is intermediate in
appearance between the profiles of HD~36861 [$\lambda$~Ori;
O8$\;$\3{((f))}, pure absorption] and HD~175754 [O8$\;$\2((f)),
P~Cygni], although the closest matches among the spectra at our
disposal are with HD~193514 [O7$\;$Ib(f)] and HD~209975 [O9.5$\;$Ib;
fig.~\ref{fig_balmer}].\footnote{We also examined the spectrum of
HD~225160; surprisingly, this O8$\;$Ib(f) star
shows much stronger emission than the O7 and O9.5 Ib stars.}
Thus while we have no exact match to the quiescent spectrum, the \Ha\
profile in that state seems to be largely unremarkable, with no
evidence of substantial excess emission relative to normal stars of
broadly comparable spectral type.

At other phases, the line-profile morphology is P-Cygni-{\em like},
but the increase in emission is not necessarily associated with an
increase in the global mass-loss rate (cf.~Sec.~\ref{sec_UV}).
Phenomenologically, the changes in the appearance of the profile can
be entirely accounted for by variable amounts of roughly gaussian
emission superimposed on a constant, underlying quiescent-state
spectrum (Fig.~\ref{fig_balmer}).  We have characterized the mean
velocity displacement and width of this excess emission by gaussian
fits; results are incorporated in Fig.~\ref{fig_Ha}.  Significant
phase dependence is evident for neither mean velocity nor fwhm, which
average +7~\kms\ and 271~\kms, respectively (standard deviations 10
and 16~\kms, commensurate with likely observational uncertainties);
the width of the excess emission is much less than both the
stellar-wind terminal velocity (Sec.~\ref{sec_HaMdot}), and the
`windy' \Ha\ emission normally seen in luminous O~stars.

Finally, although the \Ha\ and {\em Hipparcos} light-curves are in
phase (to within the errors), we note that the photometric variations
are too large to result from excess line emission alone; most of the
$\sim$3--4\%\ flux change must arise from true continuum-level variations.

\section{The 538-d period:  other variable lines}
\label{sec_var}

\subsection{Balmer and \hei\ lines}

\begin{figure*}
\center{\includegraphics[scale=0.91,angle=0]{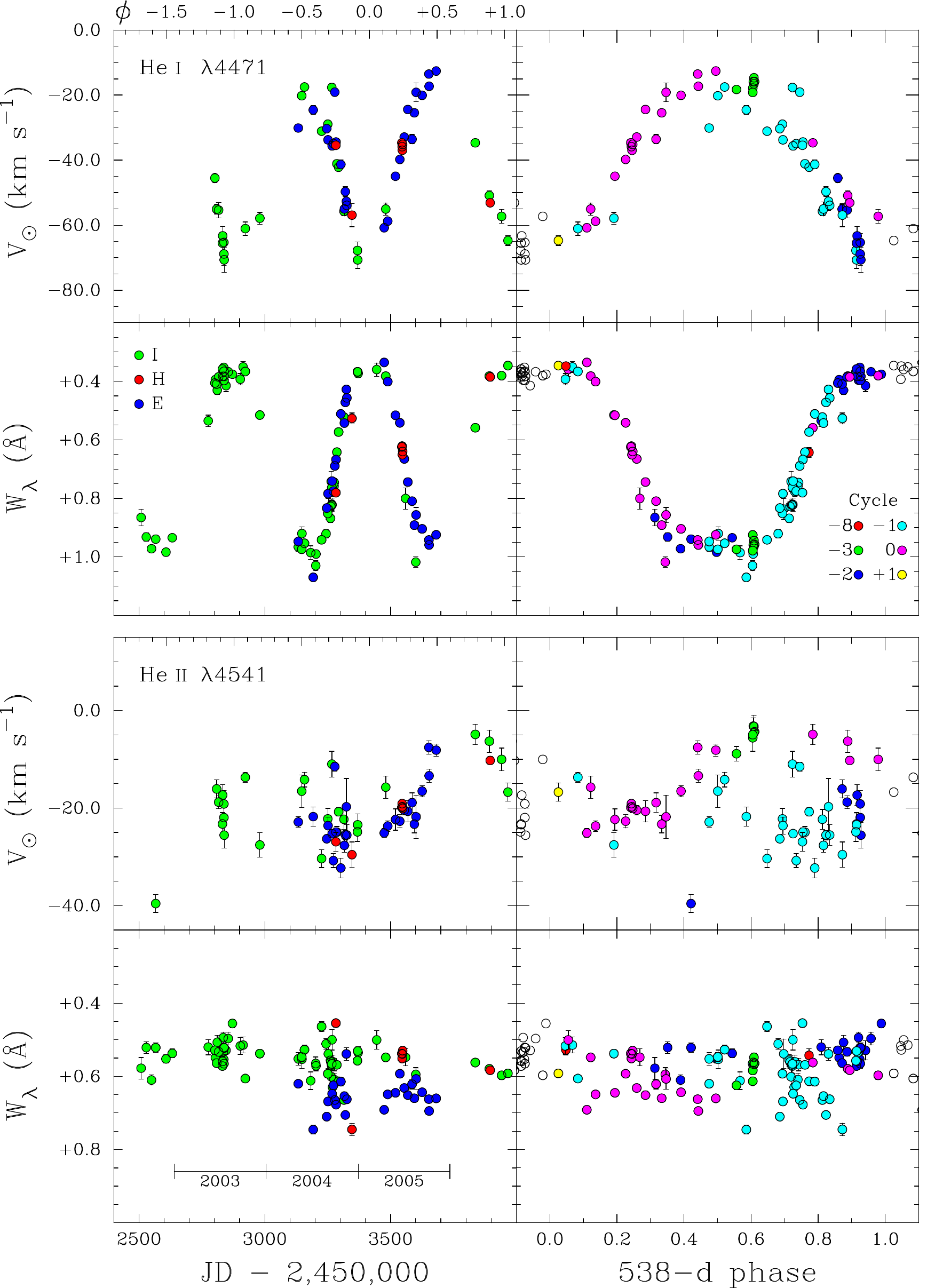}} 
\caption[]{Equivalent-width and velocity variations for \hei~4471{\AA}
and \heii~4541{\AA}.  Points with large formal measurement errors are
omitted for clarity (so that fewer velocities than line strengths are
plotted).  As in Fig.~\ref{fig_Ha}, the equivalent-width axis is inverted
(such that lower points indicate greater absorption/less emission).
Points in the left-hand panels are colour coded according to source:
E(lodie; \citealt{Naze07}), H(igh-resolution), or
I(ntermediate-dispersion). Points in the right-hand panels are
colour-coded according to cycle in the ephemeris of
eqtn.~\ref{eqn_ephem}.}
\label{fig_4471}
\end{figure*}

\begin{figure}
\center{\includegraphics[scale=0.43,angle=0]{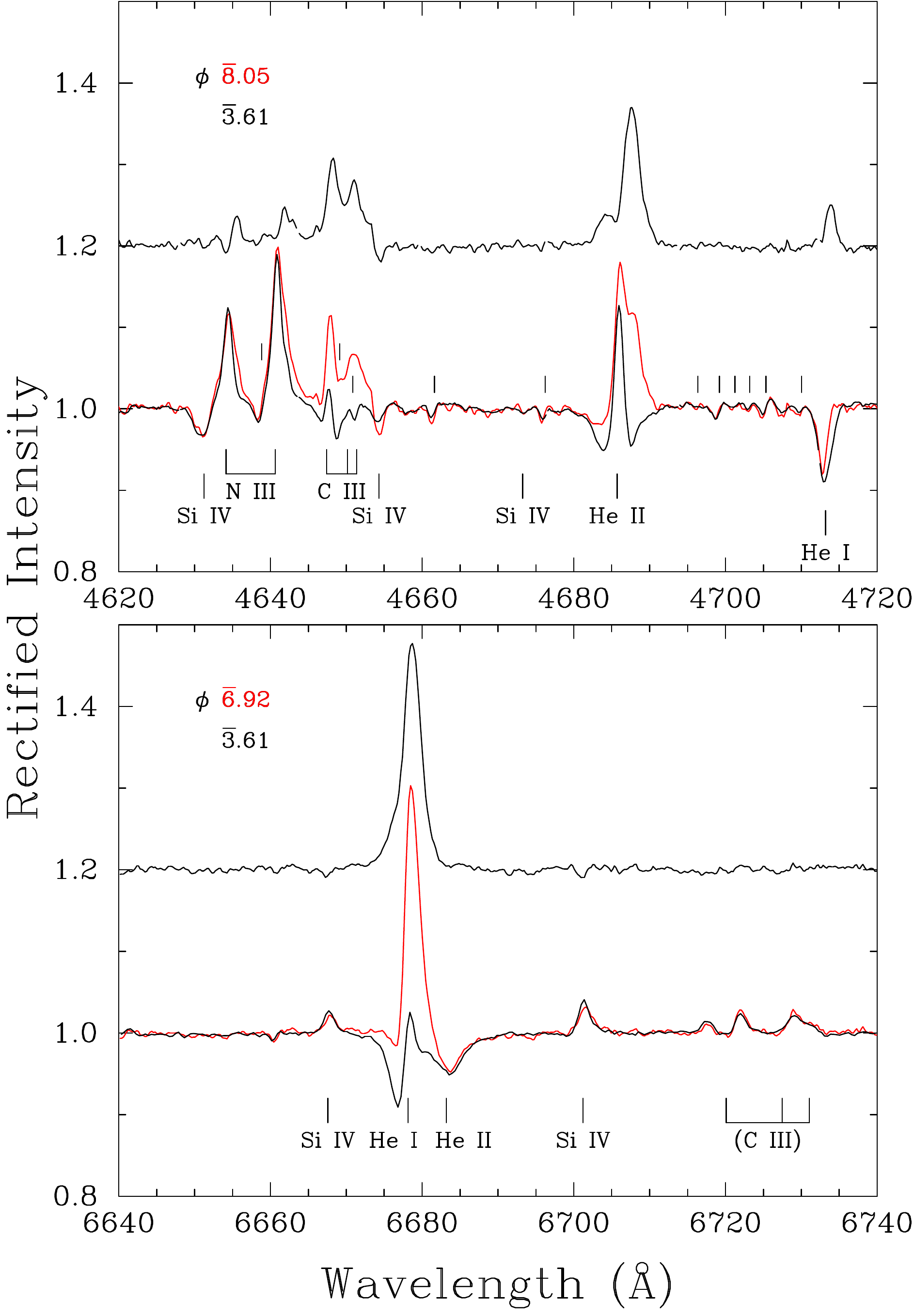}} 
\caption[]{Emission-line variability in HD~191612; spectra are
  labelled by phase in the \Ha\ ephemeris of eqtn.~\ref{eqn_ephem}.
  Data obtained near \Ha\ maximum are shown in red, those near minimum
  in black; ratio spectra are shown offset by +0.2.  (All spectra were
  obtained near orbital zero; Sec.~\ref{sec_rvv}.)  \newline
  \textit{Upper panel:} the \niii~4640{\AA}, \ciii~4650{\AA},
  \heii~4686{\AA} emission-line complex; unlabelled tick marks
  indicate wavelengths of \oii\ lines.  Note also the infilling of the
  red wing of \hei~4713\AA.\newline \textit{Lower panel:}
  \hei/\heii~6678/6681{\AA}.  Although $\lambda$6678 is unusual among
  the \hei\ lines in showing emission throughout `quiescence', the
  qualitative nature of the variability (i.e., growth of a slightly
  redshifted, fairly narrow emission superimposed on the
  quiescent-state spectrum) is notable only for the unusually large
  relative amplitude.  The $\lambda\lambda$6717/6722/6729 lines (the
  latter probably a blend in our data, $\lambda$6728.8+6731.1:) are
  almost constant in strength; the positions appear not to be
  consistent with a standard \ciii\ identification.  }
\label{fig_EmSpec}
\end{figure}

Other Balmer lines (to at least H10)\footnote{Paschen lines from 3--8
  to 3--20 are recorded in emission in the CFHT ESPaDOnS spectra;
  these are also strongly variable, in a manner consistent with the
  \mbox{538-d} period.}  and the \hei\ lines all show large
variations, in both strength and velocity, which are reproducible on
the \mbox{538-d} period.  \hei~4471{\AA} (Fig.~\ref{fig_4471})
exemplifies the typical behaviour of the \hei\ lines; most remain in
absorption at all phases, although \hei~$\lambda\lambda$5876, 7065,
7281, as well as \Ha\ and \Hb, show strong P-Cygni-like emission at
maximum. While most lines are in absorption during quiescent phases
(even \Ha, as illustrated with other Balmer lines in
Fig.~\ref{fig_balmer}), \hei~6678{\AA} is exceptional in retaining an
emission core (Fig.~\ref{fig_EmSpec}).

The basic phenomenology of the variability is, evidently, infilling of
a near-normal underlying absorption profile by slightly
($\sim$20--30~\kms) redshifted, almost symmetrical emission (cf.\
Fig.~\ref{fig_balmer}).  This interpretation is consistent both with
direct inspection of line profiles, and with the phase dependence of
variability; all lines share a common velocity around $\phi_\alpha
\simeq0.5$, becoming increasingly blueshifted with decreasing line
strength (i.e., increasing emission infill).  Moreover, the amplitude
of radial-velocity variation correlates with the amplitude of
line-strength variation.

This phenomenology appears to be applicable to almost all variable
features, embracing not only the hydrogen and \hei\ lines (including
$\lambda$6678), but also the signature Of?p \ciii~4650{\AA}
transitions (Fig.~\ref{fig_EmSpec};  the adjacent \niii\ lines are
much less, if at all, variable, and so are evidently dominated by
`normal', photospheric, Of emission).
\heii~$\lambda$4686 shows an
exceptional pattern of variability; narrow emission is present even at
quiescent phases, and while the overall emission-line strength
increases near phase zero, only small changes occur at the position of
this narrow emission (Fig.~\ref{fig_EmSpec}), possibly because the
emission is already optically thick here.

\subsection{UV P-Cygni  profiles}
\label{sec_UV}

\begin{figure}
\center{\includegraphics[scale=0.34,angle=-90]{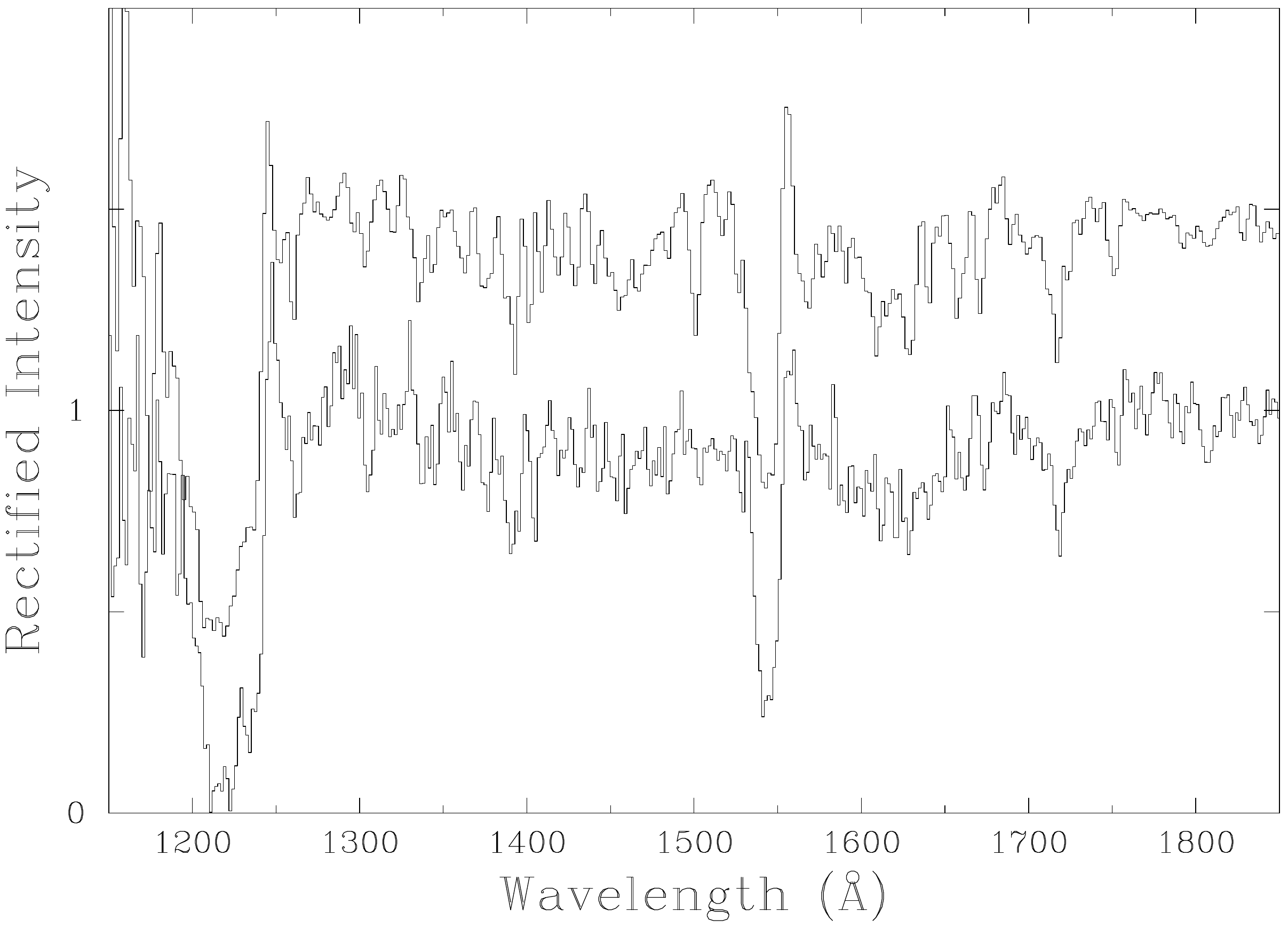}} 
\caption[]{IUE spectra of HD~191612.
Lower spectrum, low-resolution, $\phi_\alpha=\overline{14}.04$; upper
spectrum (offset by 0.5 continuum units), high-resolution spectrum,
$\phi_\alpha=\overline{9}.74$, smoothed and binned to match the
low-resolution data.}
\label{fig_IUE}
\end{figure}

\citet{Walborn03} presented a
high-resolution International Ultraviolet Explorer (IUE) spectrum
obtained on
1992 Dec~19 (JD~2,448,975.9, $\phi_\alpha=\overline{9}.74$, $R
\simeq 10^4$, $\lambda\lambda \sim 1200{\mbox{--}}1900$\AA).   
There are no other high-dispersion UV spectra
available, but the IUE archive contains
a low-resolution, short-wavelength spectrum
 (1984 July~25, JD~2,445,907.0,
$\phi_\alpha=\overline{14}.04$, $R \simeq 300$).  After smoothing and binning
the high-resolution spectrum 
to render it comparable to the low-resolution one
(Fig.~\ref{fig_IUE}), there is no
evidence for {\em large} changes in the resonance-line profiles of
\nv~$\lambda$1240, \siiv~$\lambda$1400 and \civ~$\lambda$1550
(although we cannot rule out variations at the moderate level observed
in the oblique rotator $\theta^1$~Ori~C; \citealt{Walborn94}).  This
encourages the view that the \Ha\ variability results from the
changing visibility of a constrained plasma, rather than a large-scale
global change in outflow characteristics -- e.g., if the variability
were a consequence of changes in mass-loss rate, this would be
expected to have a clear signature in the UV P-Cygni profiles.

\section{The constant lines}

\begin{figure*}
\center{\includegraphics[scale=0.91,angle=0]{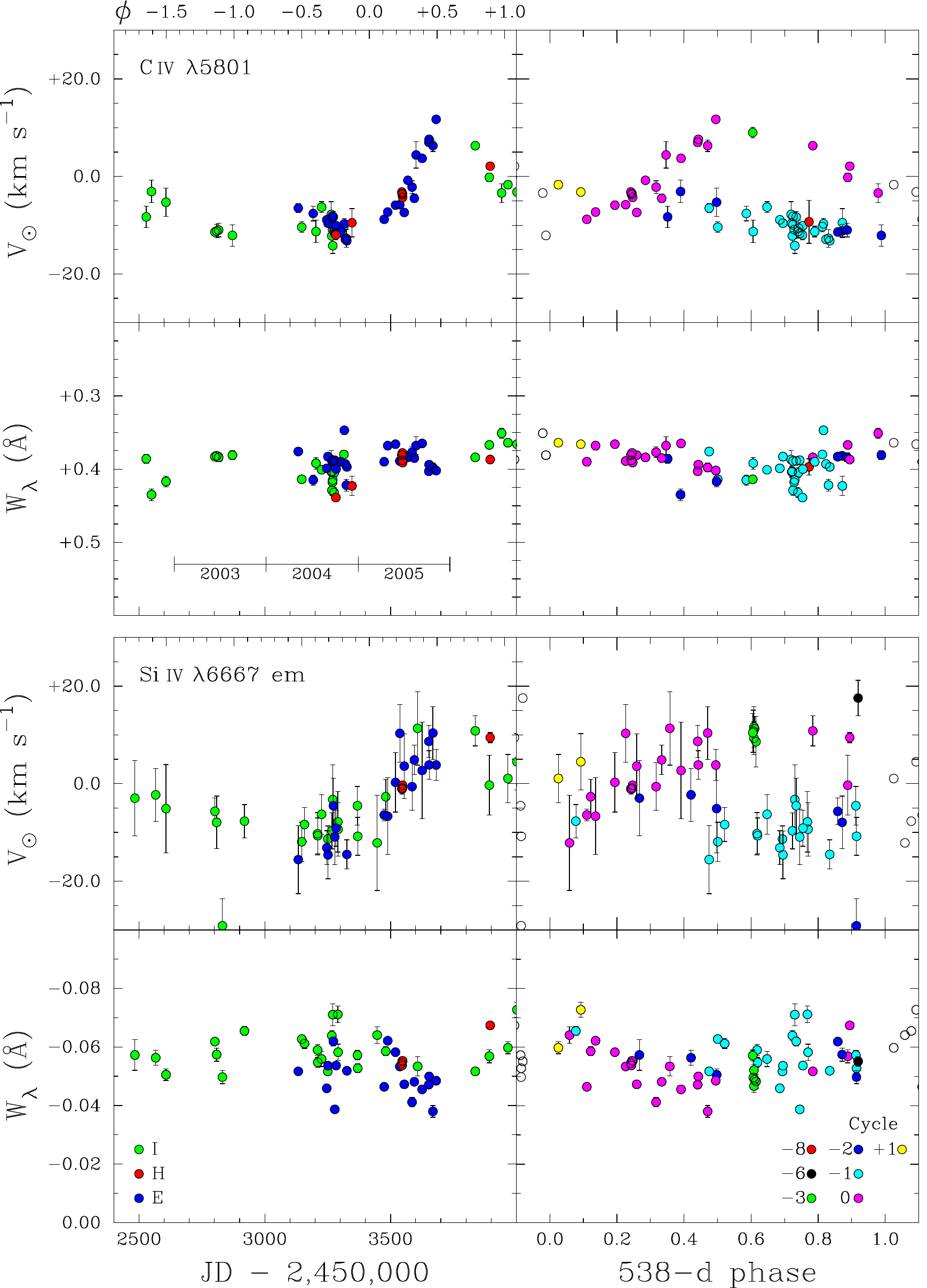}} 
\caption[]{Equivalent-width and velocity variations for the
\civ~5801{\AA} absorption and \siiv~6667{\AA} emission lines;
other details are as
for Fig.\ref{fig_4471}.}
\label{fig_5800}
\end{figure*}

In contrast to the hydrogen and \hei\ lines, the absorption lines of
metals and of \heii, together with many selective emission lines,
show only small changes, at most, in line strength.   For simplicity,
we label such lines as `constant';  if there is any line-strength
variability, it is at a very low level.

\subsection{Absorption lines}
\label{sec:abslines}

The \civ~5801, 5812{\AA} doublet exemplifies the behaviour of the
constant absorption lines.   This doublet is particularly well-suited to
measurement both on astrophysical grounds (the lines are very
symmetrical, and are formed deep in the atmosphere, and hence are less
likely to be contaminated by `windy' emission than many other
transitions; cf., e.g., \citealt{Fullerton96}), and observationally (the
nearby diffuse interstellar bands at 5778/5780/5797{\AA} are of
similar strength to the \civ\ lines, and provide a very useful
zero-point calibration for the wavelength scale, allowing rather
precise differential velocities to be obtained even from
intermediate-dispersion data of unexceptional quality).

The \civ\ measurements are presented in Fig.~\ref{fig_5800}, and
illustrate the typical behaviour of essentially constant line strength
coupled with small-amplitude radial-velocity variations.  The steady
increase in velocity between JDs $\sim$2,453,500--700 discussed by
\citet{Naze07} clearly does not repeat on the \mbox{538-d} period.
Most metal absorption lines strong enough to be measured consistently
in most spectra (e.g., \niii~$\lambda\lambda$4511--4534), as well as
the \heii\ absorption lines (e.g., $\lambda\lambda$4200, 4541,
5411) follow essentially the same behaviour as the archetypal \civ\
lines, as illustrated in Fig.~\ref{fig_4471} by data for
\heii~$\lambda$4541.

\subsection{Selective emission lines}

The high-quality UES and CFHT spectra, in particular, reveal a rich
spectrum of selective emission lines (cf.\ \citealt{Walborn01}).  In
general, the weakness of the lines precludes detailed scrutiny of
their phase dependence in the remaining data, 
but comparison of quiescent and emission-line
phases in these echellograms establishes that most of the emission
features\footnote{Including \cii~6578, 6583{\AA};
\ciii~5696{\AA};
\nii~5001/5005, 5667--5680--5686, 6482, 6610{\AA};
\niii~5321, 5327{\AA};
\siiii(?)~4905.6{\AA};
\siiv~6667, 6701{\AA}; 
\siv~4486, 4504{\AA}; and
the unidentified emission lines first noted by
\citet{Underhill95} at 
6717.5, 6722.2, 6729.1 (probably a blend in our
data, 6728.8+6731.1:), 6744.4{\AA}.}
exhibit essentially the same
behaviour as the 
\civ\ lines (i.e., near-constant line strength and small-amplitude
radial-velocity variations).
Only the stronger lines can be consistently measured, but results for
\siiv~6667{\AA}
shown in in Fig.~\ref{fig_5800} illustrate the general accord between
the behaviours of the selective emissions and the `constant'
absorption lines.

The extensive far-red coverage of the high-quality CFHT ESPaDOnS
spectra allows many other weak emission lines to be recognized; these appear
also to show little or no variability in line strength
(at least, between the two epochs
sampled, at \mbox{538-d} phases 0.24 and 0.89; the `variable' lines discussed 
in Section~\ref{sec_var}
change substantially between these epochs).  
Most of these features
are previously unreported in O-star spectra, and currently lack persuasive
identifications.  Measured wavelengths in
\AA\ (and
possible identifications, with multiplet numbers from
\citealt{Moore45}) are, for the stronger features:
5739.8;
6394.8,  
6467.1, 
6478.7 (\niii~14?  good wavelength matches but not all
multiplet members present);
6482.3;
7002.8;
7037.2 (\ciii~6.01);
7306.9;
7455.2;
7515.7;
8019.2 (\niii~26);
8103.0 (\siiii~37);
8196.6 (\ciii~43);
8251.0;
8265.7;
8268.9;
8286.8;
9705.5,
9715.4 (both strong, broad;  \ciii~2.01).

\subsection{Radial-velocity variations \& spectroscopic orbit}
\label{sec_rvv}

\begin{figure}
\center{\includegraphics[scale=0.35,angle=-90]{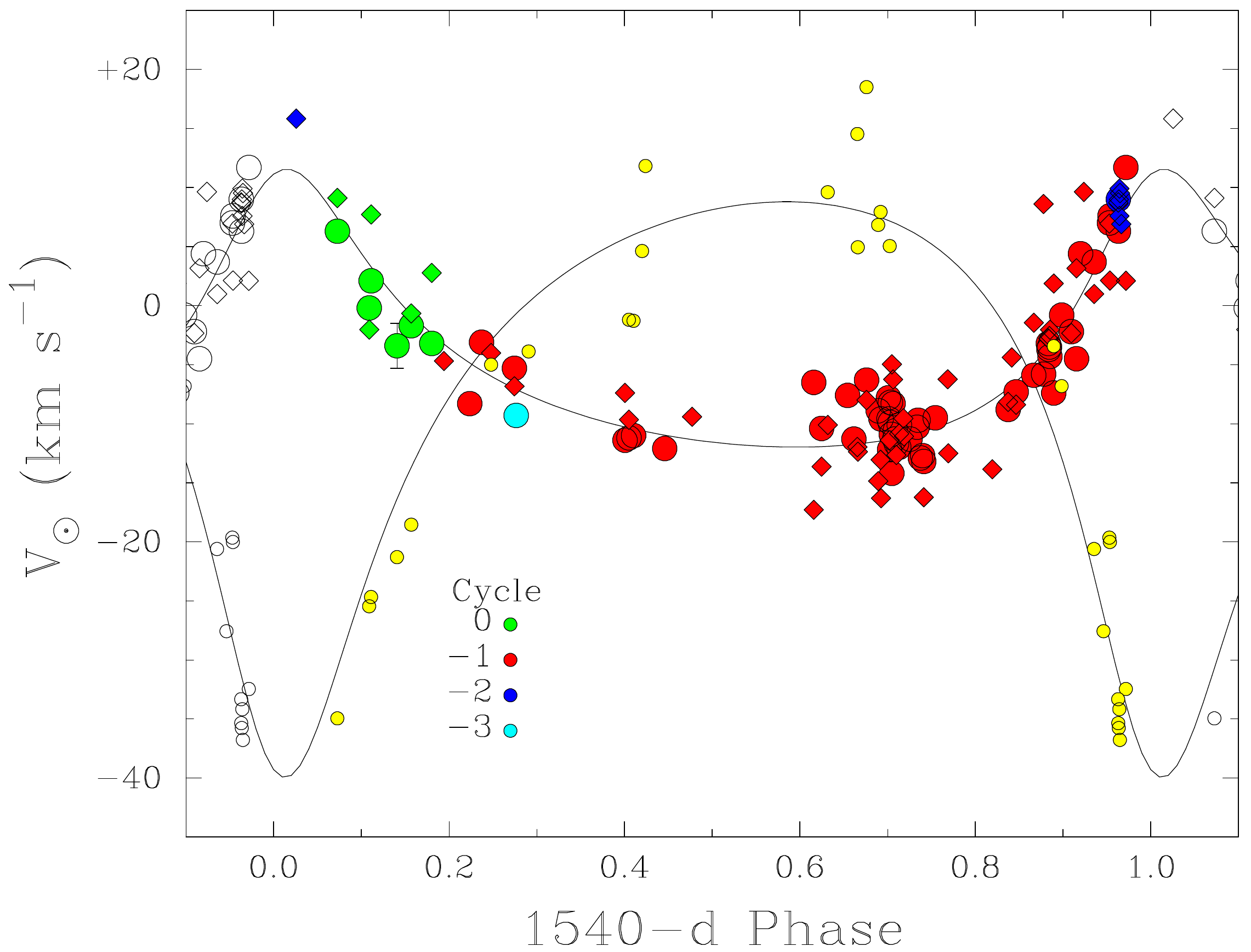}} 
\caption[]{Radial-velocity measurements for \civ~5801{\AA} (large circles),
\siiv~6667{\AA} (diamonds), and \oii\ lines (small yellow circles), plotted
with the orbital solution of
Table~\ref{tab_orbit}.  The \civ\ and \siiv\ points are colour coded
according to cycle count in the ephemeris of Table~\ref{tab_orbit},
and for display purposes the \oii\ velocities have been adjusted by
$-14.3$~\kms\ to bring them to the same $\gamma$
velocity as the primary (see Sec.~\ref{sec_rvv}).
}
\label{fig_orbit}
\end{figure}

\begin{figure}
\center{\includegraphics[scale=0.33,angle=-90]{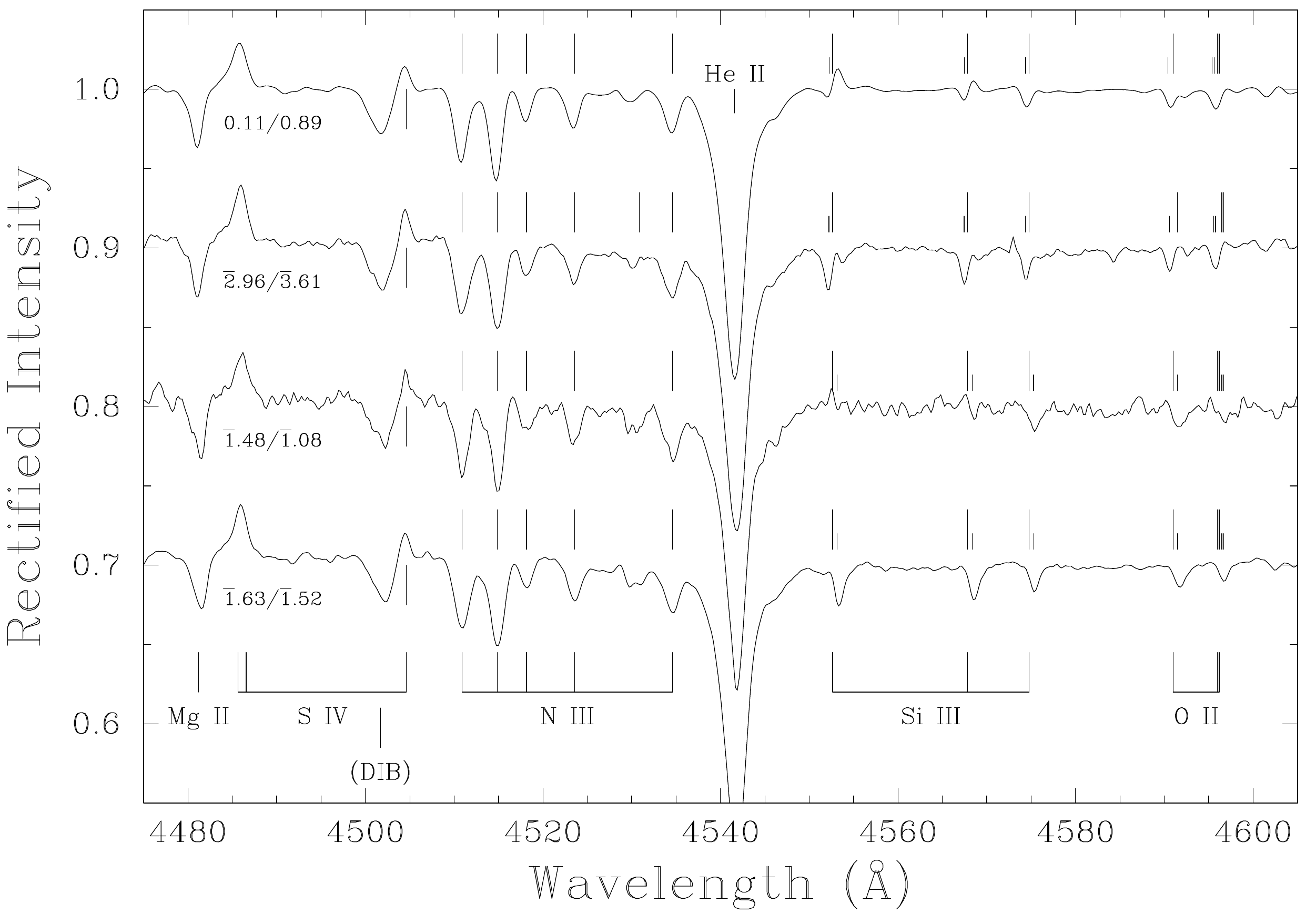}} 
\caption[]{Signatures of the
secondary's spectrum.   Data have been
lightly smoothed, rebinned, and renormalized to facilitate comparison
between observations, which are labelled by orbital/H$\alpha$ phases
(Sec.~\ref{sec_rvv}/\ref{sec_HaP}).
Spectra have been shifted to bring the primary to zero velocity; long
tickmarks indicate rest wavelengths of the identified lines.
In this reference frame the \heii, \niii\ and
\siv\ lines show no significant movement (i.e., arise
principally in the primary's spectrum).   Short tickmarks show the
expected positions of the \oii\ and \siiii\ lines in the secondary
spectrum, according to the orbital solution given in Table~\ref{tab_orbit}.
}
\label{fig_Spec1}
\end{figure}

\begin{table}
\caption[]{Orbital solution.  The main orbital parameters are
constrained by measurements of \civ~5801{\AA},
\siiv~6667{\AA} in the primary spectrum; $K_2$ is established from \oii\
lines 
in the secondary spectrum.}
\begin{tabular}{lrcll}
$\gamma$      & $-5.19$   &$\pm$&  0.36  & \kms\ \\
$K_1$         &  11.77         &&  0.84  & \kms\ \\
$e$           &  0.438         &&  0.038 \\
$\omega$      &  344.7         &&  6.5   & $^\circ$\\
$P_{\rm orb}$ &  1542          &&  14    &d \\
$T_0$         &  JD 2453720    &&  20    \\
$f(m)$        &  0.190         &&  0.042 &{\msun} \\
$a_1\sin{i}$  &  322           &&  24    &\rsun\ \\
rms residual & \multicolumn{4}{l}{(weight 1, \civ) 2.2~\kms\ }\\
\\
$K_2$      & 24.4&&1.4 & \kms\ \\
$f(m)$        &  1.68          &&  0.29  &{\msun} \\
$a_2\sin{i}$  &  667           &&  38    &\rsun\ \\
$q = M_2/M_1$&        0.483 &&0.044\\
rms residual & \multicolumn{4}{l}{(\oii) 5.1~\kms\ }\\

\end{tabular}
\label{tab_orbit}
\end{table}

The `constant' lines show significant, systematic radial-velocity
variations\footnote{The narrow core of \heii~$\lambda$4686 emission,
  present throughout quiescence, tracks the motion of the `constant'
  absorption lines and selective emission lines.}  which do not repeat
with the \mbox{538-d} period (Figs.~\ref{fig_4471} and
\ref{fig_5800}).  These velocity variations cannot reflect
photospheric motion about a static centre of mass (the implied change
in radius is as great as $\sim$250\rsun, or $\sim$17\rstar).  In
principle, they could be attributed to changes in the depth of line
formation in the accelerating outflow, through density changes in or
near the trans-sonic region resulting from stochastic changes in the
stellar-wind mass loss rate. However, the lack of changes in \Ha\ in
the mean quiescent spectra around phases $\overline{1}.5$ and 0.5
(corresponding to negative- and positive-velocity \civ\ states) does
not encourage confidence in this interpretation, and we have already
argued that the \civ\ lines are formed relatively deep in the subsonic
atmosphere.

We therefore consider the possibility of orbital motion on periods
other than that of the major spectroscopic variability.  We
concentrate on measurements of the \civ~5800{\AA} doublet and the
$\sim$6700{\AA} emission-line complex (\siiv\ + unidentified), which
yield mutually consistent results.  Both features have velocity
zero-points tied in to nearby, moderately strong diffuse interstellar
bands (we adopted DIB wavelengths as measured in the CFHT spectra),
and therefore record the velocity variations more precisely and
accurately than other lines that lack this advantage.  Results are
given in Table~\ref{tab_rvs}.

Fortuitously, the dataset includes good-quality observations showing
that relatively large positive velocities occurred in \mbox{538-d}
cycles $-3$ and $-6$, as well as at $\phi_\alpha \sim 0.5$, hinting at
$P_{\rm orb} \simeq 3P_\alpha$.  Trial orbital solutions with $P_{\rm
  orb} \simeq 1500$--1600d yielded residuals that are satisfactorily
small, but none the less slightly larger than the formal errors
returned by the gaussian fits to lines used to measure velocities.
Since external errors evidently dominate (\civ) or match (\siiv) the
formal uncertainties, we weighted all \civ\ measurements and all
\siiv\ measurements equally, but with the \siiv\ measurements assigned
$\sim$1/3 weight to reflect their greater residuals (and adjusted by
$-2.1$~\kms\ to bring them to the same $\gamma$ velocity as the \civ\
measurements).

The resulting orbital solution is summarized in Table~\ref{tab_orbit},
and illustrated in Fig.~\ref{fig_orbit}.   While this solution has
not yet been subject to the acid test of predictive power, its success
in reproducing observations at three separate periastron passages,
with small residuals that are consistent with realistic observational
uncertainties, encourages the view that the solution does characterize a
true binary orbit.

\subsubsection{Secondary orbit}

Careful examination of weak lines of low ionization stages in the best
data shows velocity displacements in antiphase with the stronger
lines, bolstering the interpretation of the \mbox{1540-d} velocity
variations as being orbital in origin.  The spectroscopic fingerprint
of the secondary is weak, but its velocity can be reasonably well
quantified by simultaneous gaussian fits to a selection of a
half-dozen unblended, relatively strong \oii\ absorption lines
($\lambda\lambda$4300--4700\AA, central depths 2--3\%\ below
continuum), constrained to share the same velocity shift; results are
included in Table~\ref{tab_rvs}.

Although we don't have the benefit of well-defined interstellar
features to provide a velocity reference for most observations
encompassing the \oii-line spectral region, we can exploit the fact
that the centrally placed \heii~4541{\AA} line follows the motion of
the \civ\ and \siiv\ features used to determine the primary's orbit
(Sec.~\ref{sec:abslines}).  By measuring \oii\ velocity {\em
  differences} with respect to $\lambda$4541, and correcting for
computed motion of the primary, we can in effect remove
spectrum-to-spectrum velocity zero-point offsets.  (Using directly
observed velocities gives essentially identical results, but with
somewhat larger scatter.)

An orbital solution of the resulting \oii\ velocities gives $\gamma$
and $K_2$ (with all other parameters fixed at the primary-orbit
values\footnote{For completeness we report that $\gamma$ velocity for
  the \oii\ lines is, formally, $+9.1{\pm}1.0$~\kms.  Although small
  differences in $\gamma$ velocities for different lines would not be
  unusual in an O-type star, it should be noted that our $\gamma$
  velocities for the \civ\ doublet and the \siiv/$\lambda$6700 complex
  depend on {\em measured}, (not laboratory) wavelengths for both the
  reference DIBs and the unidentified $\lambda$6700 emissions, while
  the weakness of the \oii\ lines means that small zero-point shifts
  arising from, e.g., unrecognised blends would not be
  surprising.\label{foot_rvv}}); the slope of \heii\ vs.\
\heii$-$\oii\ velocities gives an entirely consistent mass ratio.
Results are incorporated into Table~\ref{tab_orbit}.

\subsubsection{Relationship between the 538- and 1540-d periods?}
\label{sec:relate}

The orbital period is $\sim{3}\times{538}$d, which has suggested to
colleagues the possibility of some sort of resonance.  However, a
solution with $P_{\rm orb} \equiv 3P_{\alpha}$ gives a significantly
poorer fit than one with $P_{\rm orb}$ allowed as a free
parameter.\footnote{An $F$~test yields a probability of $\ll$1\%\ that
$\chi^2$ for the fixed-period fit is no poorer than that for the
solution with $P_{\rm orb}$ free; i.e., formally rules out $P_{\rm
orb} \equiv 3P_{\alpha}$ with $>99$\%\ confidence.} Thus although it
seems to be quite securely established that HD~191612 is a long-period
binary, it appears that, with $P_{\rm orb} = 2.87P_\alpha$, the binary
orbit has no important role in the large-amplitude spectroscopic
variability.

There is, however, an interplay between the periods in a limited
observational sense, explaining otherwise complex behaviour seen in
some lines.  Fig.~\ref{fig_Spec1} shows spectra obtained near the
extremes of both the orbital motion and the spectroscopic variability.
In addition to illustrating the small orbital velocity amplitudes
(less than the line widths) the quiescent-state spectra clearly
indicate that absorption in the \siiii~$\lambda\lambda$4552, 4568,
4575 triplet tracks the secondary spectrum, on the orbital period.
The {\em primary} spectrum has emission in these lines which varies on
the \mbox{538-d} period, and which shows a strong decline in strength
going from $\lambda$4552 to $\lambda$4575 (as is observed in some
LBV/WN11 spectra; e.g., \mbox{He 3-519}, \citealt{WalbornF00}).

The interplay of these two cycles means that the lines can appear
purely in absorption (throughout quiescent \mbox{538-d} phases), but
in other states the behaviour is more complex.  When the primary and
secondary spectra are at or near maximum velocity separation, a
P-Cyg-\textit{like} appearance results in $\lambda\lambda$4552, 4568,
with absorption in $\lambda$4575 (as in the first [top] spectrum in
Fig.~\ref{fig_Spec1}); emission phases at smaller velocity separations
can result in apparent disappearance, or nulling, of $\lambda$4552
(third spectrum in Fig.~\ref{fig_Spec1}).

A similar effect is detectable in the \nii\ spectrum, in particular
the 5001/5005{\AA} lines, where primary emission and secondary
absorption clearly move in antiphase; and, possibly, in the
$\lambda\lambda$4414.9, 4417.0 \oii\ lines.

\subsection{System characteristics}

\begin{figure}
\center{\includegraphics[scale=0.35,angle=-90]{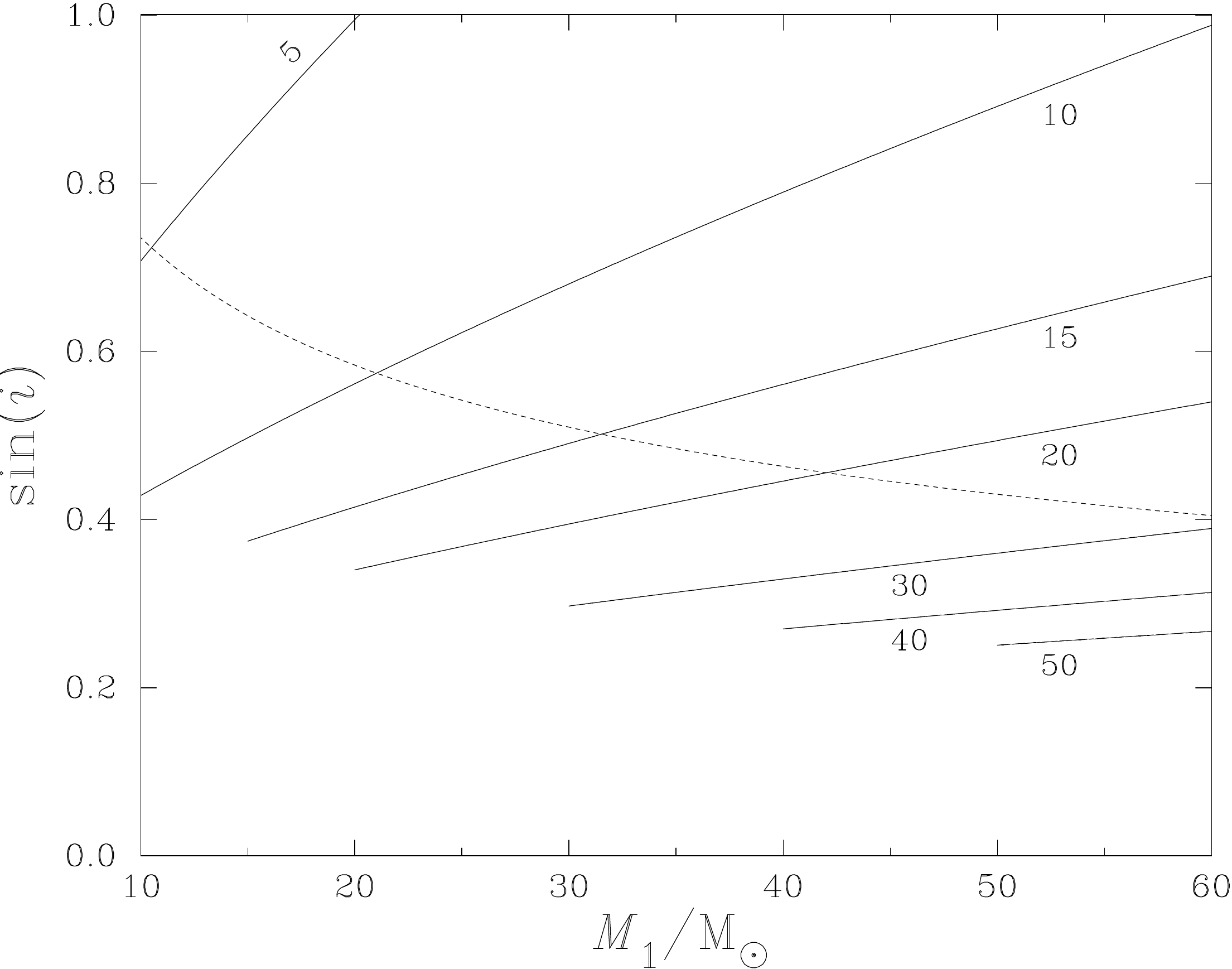}} 
\caption[]{Constraints on the system mass resulting from the orbital
solution of Table~\ref{tab_orbit}.  Solid lines are loci of constant
secondary mass, labelled in solar masses, implied by the mass
function (assuming $M_2 \le M_1$);  the dotted line shows the estimated mass ratio.
}
\label{fig_m1sini}
\end{figure}

Mass constraints implied by the orbital solution are illustrated in
Fig.~\ref{fig_m1sini}.  We have no constraint on the orbital
inclination, but the masses are entirely consistent with a system
comprising a $\sim$30-\msun, late-O giant accompanied by a
$\sim$15-\msun, early-B main-sequence star, viewed at an intermediate
angle ($\sin{i} \simeq 0.5$).  The projected centres-of-mass
separation at periastron is $(1-e)a\sin{i} = (555 \pm 68)\rsun$
($\sim{40}\rstar$); although this certainly allows for the possibility
of wind--wind interactions, the H$\alpha$ emission region, for example,
is surely much closer to the primary, so that it is unlikely that the
secondary plays any important role in the 538-d changes.

The strong variability of the \hei\ classification lines, together
with the small orbital velocity amplitudes (which ensure that the
components' helium lines are never resolved in the spectrum) means we
cannot estimate the spectral types separately from the spectrum; the
{\em maximum} velocity separation is less than the line widths in the
primary.  Thus even if the components were as just postulated, with an
expected $B$-band brightness ratio of $\sim$10:1, the combination of
the likely differences in spectral types and the resolution and
signal:noise of most of our spectra means it is not particularly
surprising that we don't see truly double-lined features (other than
the peculiar \siiii\ lines noted in section~\ref{sec:relate}).
Nonetheless the \oii/\siiii\ lines indicate a secondary spectral type
not earlier than B0, and not later than B2 (assuming a dwarf
luminosity class and normal composition).



\section{Other spectroscopic properties}

\subsection{Spectral classification}

Our extended dataset, and improved understanding of the spectroscopic
variability, allow us to revisit the spectral classification (that is,
the classification of the {\em spectrum} -- not the component stars).

The $\phi_\alpha = 0$ spectrum (specifically, the WHT spectrum of JD
2453941.5) yields spectral type O6.5f?pe (`O6.5' from the
\hei~4471/\heii~4541 ratio, with qualifiers `f?p' from
\ciii~$\lambda$4650 and `e' from the H$\gamma$ emission). Our
refined classification of the quiescent-state spectra is
O8fp (the weakness of \ciii\ in that state excluding Of?p). Of
course, it is now clear that these apparent variability in spectral
type does not reflect changes in any fundamental stellar properties --
which would result in large-amplitude photometric variability -- but
instead results from infilling of the \hei~$\lambda$4471
classification line by emission.

These results are in full accord with previous classifications,
including in particular \citeauthor{Walborn73}'s original blue-region
spectrum, obtained on 1972 Sept.~24 ($\phi_\alpha = \overline{22}.02$)
and also classified O6.5f?pe \citep{Walborn73} -- consistent with the
underlying `clock' keeping good time for at least a decade before the
earliest digital data.  

Luminosity classification at any state is not possible because of the
unique, peculiar profile of the fundamental \heii~4686\AA\ criterion.
The strength of \siiv~4089{\AA} relative to nearby \hei\ lines would
be consistent with luminosity class~\5, while the UV \siiv\ resonance
doublet is neither dwarf- nor supergiant-like, but is possibly
consistent with class~\3.  The physical properties established in
Section~\ref{sec_EBV} are also broadly giant-like.  However, all the
known Of?p stars are sufficiently spectroscopically peculiar that any
luminosity-class argument based on direct comparisons with normal
stars must be considered hazardous.

\subsection{Spectroscopic fine analysis}

Although the peculiarities of the spectra preclude extremely detailed
modelling, the basic photospheric parameters can none the less be
quite well constrained.  We searched for matches to the `quiescent'
spectrum of 2001 Aug~5 ($\phi_\alpha=\overline{3}.61$), using a grid
of {\sc fastwind} models \citep{Puls05} sampled at steps of 1kK and
0.2 in \Teff\ and \logg\ (with $Y = 0.1$ and $v_{\rm turb} =
15$~\kms).  We find a generally good match with observations for a
model with $\Teff = 35.0$kK, $\logg = 3.5$, in excellent agreement
with the analysis reported by \citeauthor{Walborn03}
(\citeyear{Walborn03}, conducted by AH but based on simpler models and
poorer-quality data).

No model reproduces the strength of \hei~4471{\AA} (cf.\ the general
discussion of this issue by \citealt{Repolust04}), but the \hei\
singlet lines are fully consistent with the estimated \Teff; adopting
34kK would still require $\logg=3.5$, but leads to the model \heii\ lines
becoming too weak.  The best-fit model parameters can therefore be
considered as fairly well determined, to $\sim\pm$1kK, 0.1 in \Teff,
\logg\ (although, of course, no model matches the peculiar \heii~4686,
\hei~6678{\AA} profiles).

This takes no account of the effect of the secondary spectrum, which
is expected to have somewhat stronger \hei\ lines, while the \heii\
line strengths may be underestimated by $\sim$10{\%} through dilution.
This would result in the primary being, if anything, modestly warmer
(by perhaps $\sim$1kK) than the foregoing analysis suggests.

\subsection{Reddening, radius}
\label{sec_EBV}

\begin{figure}
\center{\includegraphics[scale=0.40,angle=-90]{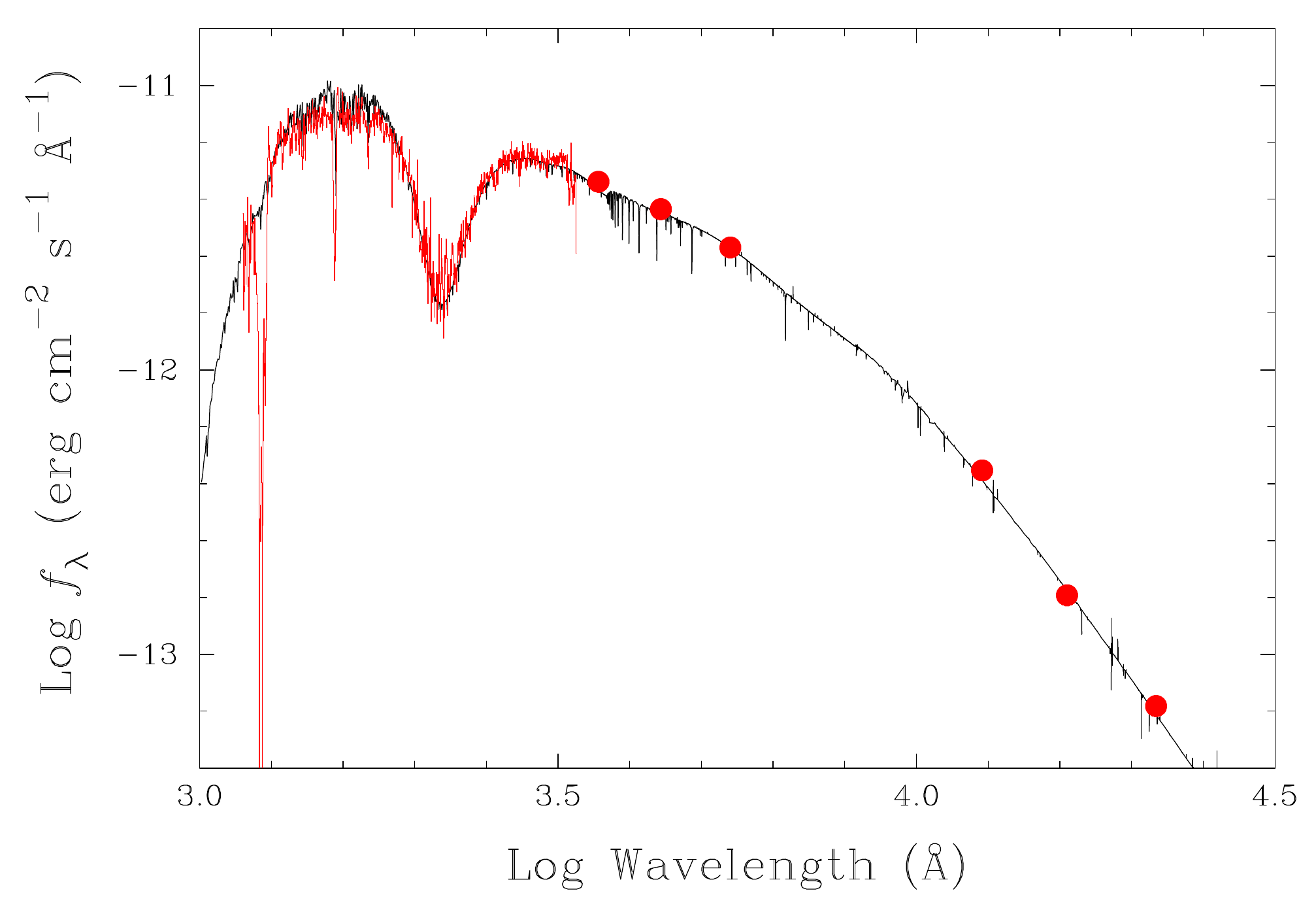}} 
\caption[]{Spectral-energy distribution for HD~191612 (red).
The weighted sum of an
{\sc Ostar2002} model with $\Teff = 35$kK, $\logg = 3.5$
and an {\sc Atlas9} model with
$\Teff = 20$kK, $\logg = 4.5$, with a $V$-band flux ratio of 9:1
and
$E(B-V) = 0.56$, is shown for comparison (black;
see Section~\ref{sec_EBV} for details).}
\label{fig_sed}
\end{figure}

To determine the reddening and angular diameter we compared a $\Teff =
35.0$kK, $\logg = 3.5$ {\sc Ostar2002} model \citep{Lanz03} with
archival low-resolution IUE spectrophotometry and optical \& {\sc
2mass} photometry.  We estimate $(R_{\rm eff}/\rsun)/(D/\mbox{kpc}) = 6.7
\pm 0.05$ and $E(B-V) = 0.56 \pm 0.03$ [using a \citealt{Cardelli89}
reddening law with $R_V \equiv A(V)/E(B-V) =3.1$]; here $R_{\rm eff}$ is
an `effective' radius characterizing the emitting surfaces.  The
match between the model and observations, though not perfect, is
reasonably good, and there is no evidence of any IR excess out to
2$\mu$m.  The fit is slightly improved by allowing for the
contribution of a cooler secondary, such as a $\sim$20kK secondary
contributing $\sim$10\%\ of the light at $V$, and this {\em ad hoc}
model is illustrated in Fig.~\ref{fig_sed}.

The reddening and general properties are consistent with membership of
the Cyg~OB3 association, as noted by \citet{Humphreys78}; we adopt her
association distance of 2.29~kpc (see discussion in \citealt{Walborn02}
for the embedded cluster NGC~6871), whence the
primary's radius is
\[
\rstar \simeq 14.5
\sqrt{\frac{f_1}{0.9}}
\left({\frac{d}{2.3 \, \mbox{kpc}}}\right)
\left({ \frac{35\mbox{kK}}{\Teff} }\right)
10^
{+0.2
\left[{
\frac{R_V}{3.1}
\frac{E(B-V)}{0.56}
}\right]
}
\rsun,
\]
where $f_1$ is the fractional $V$-band luminosity of the primary and
the $\Teff$ term accounts for the approximate $\Teff^2$ scaling of
model-atmosphere $V$-band surface fluxes in this temperature range.
The primary then has  $\log(L/\lsun) \simeq 5.4$ and $M(V) \simeq
-5.6$ -- parameters broadly consistent with a late-O giant.

\subsection{H$\alpha$ mass-loss rate}
\label{sec_HaMdot}

In our {\sc fastwind} models the mass-loss rate is characterized by a
parameter
\[
Q = \frac {\mdot \sqrt{f_{\rm cl}}}  {\msun \mbox{yr}^{-1}} 
\left/ {
\left({   \frac{\vinf} {\mbox{\kms}}
          \frac{\rstar}{\rsun}}\right)^{1.5} }
\right.
\]
\citep{Puls05}, where $f_{\rm cl} =
\left<\rho^2\right>/\left<\rho\right>^2$ is the clumping factor; from
modelling of \Ha, we find $-\log{Q} = 12.6\mbox{--}12.7$ for the
minimum-state spectrum.  We estimate $\vinf \simeq 2700$~\kms\ (with
$\sim$10--15\%\ uncertainty) from the \civ\ resonance doublet in the
IUE spectra, whence the primary's quiescent stellar-wind mass loss
rate is $\log\left({{\mdot \sqrt{f_{\rm cl}}}}\right) \simeq
-5.8$~dex~${\msun \mbox{yr}^{-1}}$.  As for other aspects of the {\sc
  fastwind} analysis, this is in satisfactory agreement with the
corresponding result reported by \citeauthor{Walborn03}
(\citeyear{Walborn03}; $\mdot = -5.6$~dex~${\msun \mbox{yr}^{-1}}$).

\subsection{Rotation}

\begin{figure*}
\center{\includegraphics[scale=0.55,angle=-90]{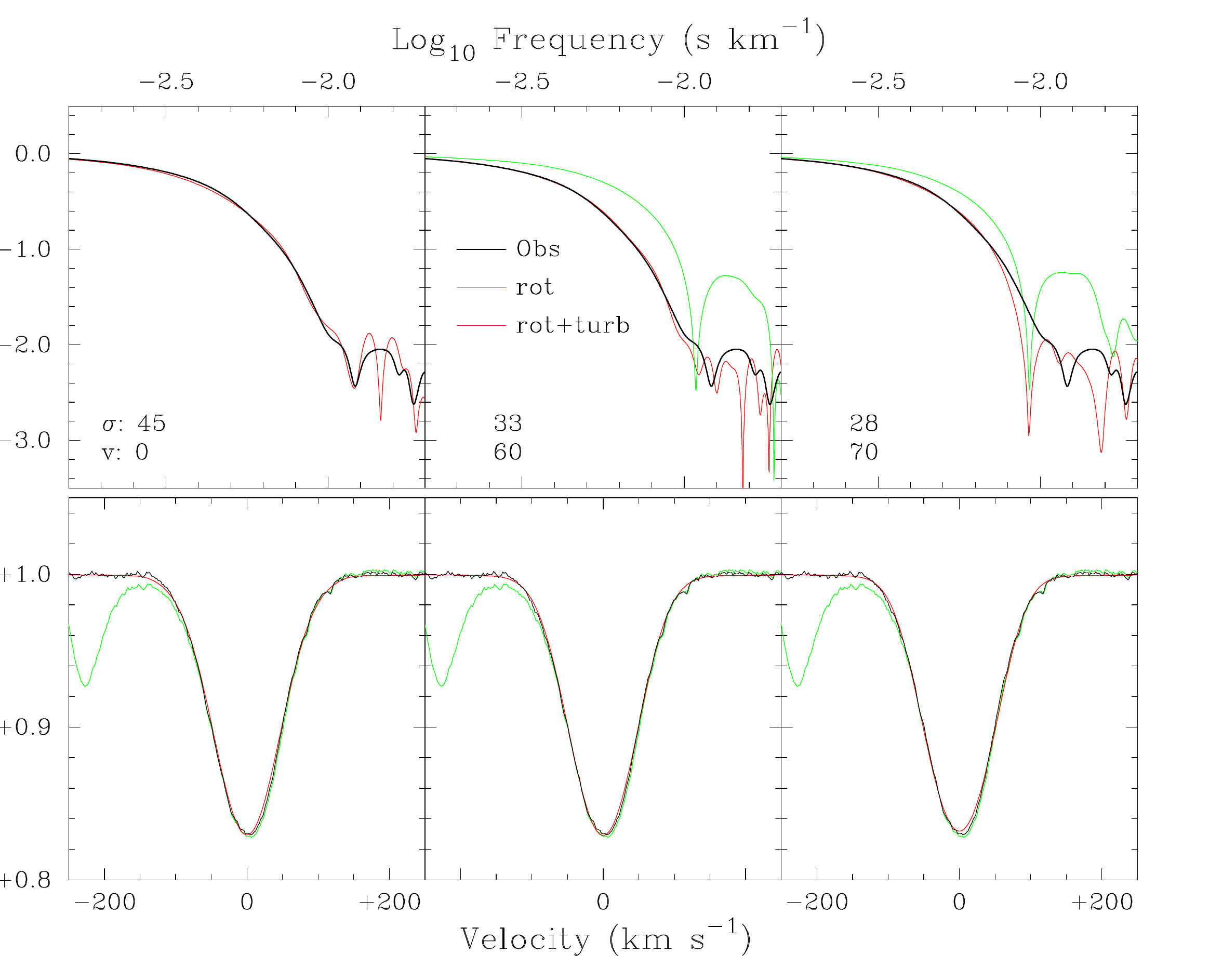}} 
\caption[]{Illustrative line-broadening models for the \civ~5801{\AA}
  line. \newline \textit{Upper panels:} results in the fourier domain.
  Solid black lines show the transform of the observed profile (i.e.,
  the transform of the black spectrum in the lower panels); red lines
  show the transforms of synthetic profiles computed with isotropic
  gaussian macroturbulence (with dispersion $\sigma$, in \kms),
  rotational broadening (with $\vesini = \mbox{v}$), and gaussian
  noise; and green lines show the transforms for rotational broadening
  alone. Any model with $\vesini \la 60$~\kms\ (and appropriate
  macroturbulence) is acceptable.\newline \textit{Lower panels:}
  results in the wavelength domain.  The observed spectrum is shown in
  each panel in green (the directly observed profile) and black (the
  version of the profile used in the analysis, after re-rectification
  to take out the diffuse interstellar band).  The synthetic profiles,
  shown in red, are the inverse transforms of the models shown in red
  in the upper panels.  At this scale, the synthetic profiles (which
  have been modestly scaled in intensity to match the observed line
  strength) are almost indistinguishable from each other, and from the
  observed spectrum, in the wavelength domain.  }
\label{fig_rot}
\end{figure*}

The equatorial velocity expected for a \mbox{538-d} rotation period
and $\rstar \simeq 14.5\rsun$ is $\sim$1.4~\kms, and a measurement of
\vesini\ would provide a strong test of rotational modulation.  (For
reference, the equatorial rotation velocity for synchronous rotation
is $\sim$0.5~\kms, or, for pseudosynchronous [periastron-synchronized]
rotation, $\sim$0.8~\kms.)  Most lines in the spectrum show
asymmetries or variability to some extent, but, as already discussed,
the \civ~5800{\AA} doublet shows exceptionally symmetrical profiles
with little or no evidence of contamination by emission; these
high-excitation lines are expected to form rather deep in the
atmosphere, and therefore to be more representative of the subsonic
photosphere than many other features.

\citeauthor{Donati06a} (\citeyear{Donati06a}a; see also
\citealt{Howarth03}) showed that the \civ\ profiles deviate strongly
from those expected from rotational broadening, but are well matched
by simple gaussian `turbulence', with zero rotation.  We have
attempted to set upper limits to the rotation rate by looking for
zero-amplitude nodes in the fourier transform of the profiles
\citep{Gray92} in the combined 2005 CFHT spectra $(\mbox{s:n} >
10^3)$.  We compared the results with transforms of synthetic spectra
that include rotational broadening, isotropic gaussian
macroturbulence, and gaussian noise, using an intrinsic profile from
an {\sc Ostar2002} model \citep{Lanz03}.

Illustrative results are shown in Fig.~\ref{fig_rot}.  Even for these
very high-quality results there is no sign of rotational zeroes above
the noise (which dominates at frequencies
$\gtrsim{10}^{-2}$~s$\;$km$^{-1}$).  Empirically, we find that the
observations are well matched by any model having
\[
\sqrt{
\sigma_{\rm V}^2 + (\vesini/2)^2} \simeq 45~\kms
\]
(where $\sigma_{\rm V}$ characterizes the velocity dispersion of the
macroturbulence and \vesini\ characterizes the projected equatorial
rotation velocity), for $\vesini \la 60$~\kms; larger values of
\vesini\ are ruled out by the data.  This is a disappointingly weak
upper limit given the data quality, but results from the effectiveness
of the gaussian turbulence in removing information with high spatial
frequencies from the profiles.

\section{Discussion}
\label{sec:disco}

\begin{figure}
\center{\includegraphics[scale=0.31,angle=-90]{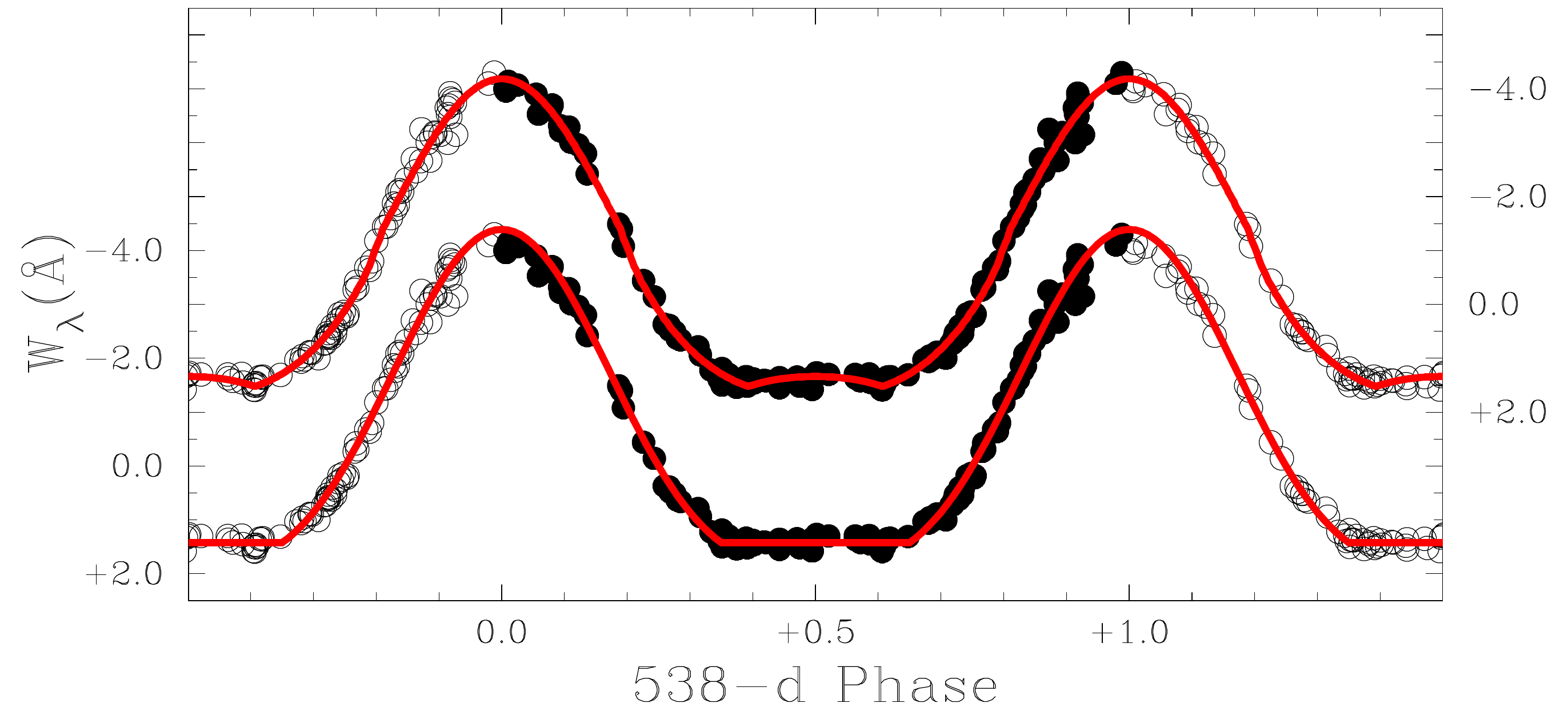}} 
\caption[]{`Toy' models compared to phased \Ha\ equivalent widths
(circles).      Lower model (left-hand axis),  a single surface
`spot'; 
upper model (right-hand axis), an optically thick disk.
See section~\ref{sec:disco} for details.
}
\label{fig_toy}
\end{figure}

The symmetry and reproducibility of, in particular, the \Ha\
\mbox{538-d} light-curve strongly suggest an origin in an essentially
geometrical process -- i.e., the changing aspect of a distinct
emission region.  This is also consistent with the absence of evidence
of associated global changes in stellar-wind properties
(Sec.~\ref{sec_UV}).  Since we have found that the variability is not
orbital, our results lend strong support to the proposal by
\citeauthor{Donati06a} (\citeyear{Donati06a}a) of rotationally
modulated line emission from a magnetically constrained plasma.

One might hope that the fairly rich emission spectrum would offer
insight as to conditions in the line-forming region, but there is a
dearth of traditional nebular diagnostics (suggesting a high-density
regime).  We can only infer some general characteristics of the
\Ha-emitting region.  First, optically thin \Ha\ emission in an {\em
  any} time-independent axisymmetric structure cannot account for the
observed 538-d behaviour; at least half the emission from any such
structure would visible at all times, but emission equivalent to the
adopted `excess' emission cannot plausibly be present during
`quiescent' phases.  At the same time, very complex geometries (such
as observed in $\tau$~Sco; \citeauthor{Donati06b}
\citeyear{Donati06b}b) are unlikely, as they would not be consistent
with the large amplitude of emission variability.

Secondly, for the distance and reddening adopted in
Section~\ref{sec_EBV}, and assuming isotropic, case-B recombination
emission, the maximum excess \Ha\ emission over the quiescent state
corresponds to
\[
\int{n_{\rm e}^2\,\mbox{d}V} \simeq 7 \times 10^{22} 
\mbox{cm}^{-6} \rstar^3.
\]
This is a lower limit to the true emission measure if
the emission is not optically thin.

If we adopt the hypothesis that the geometry of the line-emitting
region is determined by the magnetic field, then the simplest
acceptable magnetic-field geometry is a misaligned, centred dipole --
known to provide a reasonable approximation to $\theta^1$~Ori~C, the
prototype magnetic O~star \citep{Stahl96, Donati02, Wade06}.
\citet{Babel97} argue that, for such a geometry, stellar-wind material
from the magnetic polar and temperate regions will be deflected along
the field lines towards the magnetic equator, where the colliding
flows shock to create a high-temperature, X-ray emitting plasma.  The
plasma cools to form a dense disc which gives rise to optical
emission. MHD calculations support this general outline, albeit with a
number of refinements (e.g., \citealt{udDoula02, Gagne05, udDoula06}).
 
Since this model is at least consistent with the direct magnetic-field
measurements of HD~191612 (\citeauthor{Donati06a}
\citeyear{Donati06a}a), for heuristic purposes we explore schematic
`toy' models of \Ha\ emission from a centred, tilted, geometrically
thin disk, in which the relative emission is simply related to the
projected area (taking into account limb darkening and occultation by
the star).  Although this is a physics-free, geometrical model, we can
speculate that it may characterize a more realistic scenario; e.g.,
for a disk-like emission region, the `limb darkening' may actually
correspond to increasing \Ha\ optical depth as the line of sight
approaches the plane of the disk.

We find that this simple model can provide a reasonable match to the
\Ha\ variability, provided that (i) the sum of the inclination of the
rotational axis to the line of sight and the angle between that axis
and the magnetic axis is close to 90$^\circ$, and (ii) moderately
strong limb-darkening is present (so that the emission is low at all
rotational phases when the line of sight is close to the disk plane).
Moreover, the width:thickness ratio must be reasonably large, to
account for the large amplitude of emission-line variability.

A specific model, selected through a genetic-algorithm minimization,
is shown in Fig.~\ref{fig_toy}; its parameters are inner radius
$R_{\rm in} = 2.00R_*$, outer radius $R_{\rm out} = 2.13R_*$, axial
inclination $i = 68^\circ$, magnetic-axis offset $\alpha = 27^\circ$,
and linear limb-darkening coefficient $u = 0.7$.  These parameters
should be regarded only as illustrative, as many other combinations
give closely similar fits (e.g., $1.55R_*$, $1.76R_*$, 41$^\circ$,
56$^\circ$, 0.6 gives results almost indistinguishable from those in
Fig.~\ref{fig_toy}).  The radii, in particular, are only very weakly
constrained from the \Ha\ light-curve, because the total emission is
fixed by an arbitrary scaling factor (parametrizing the surface
emissivity).

[In principle, the continuum photometry could help fix the radii, but in
practice it only weakly constrains the model, because the photometric
amplitude is so small, the noise is relatively high, and the number of
free parameters is large.  The only firm conclusion is that a very
large disk, very close to the star, and optically thick in the
continuum, is not allowed.  The broad-band photometry suggests that
any disk cannot contribute more than $\sim$3--4\%\ of the
visible-region broad-band ($H_P$) flux -- that is, the implied
equivalent width of the \Ha\ emission referred to disk continuum is as
much as $\sim$150{\AA}.]

The toy model succeeds in reproducing the \Ha\ light-curve, but it
would clearly be rash to put too much weight on this success.  Not
only does the physical model which inspired it encounter difficulties
in accounting for the details of the observed X-ray emission
\citep{Naze07}, but also the disk model is certainly not unique.  To
illustrate this, we have also considered a minimalist model of \Ha\
emission from a single surface spot,\footnote{Though it lacks any
  compelling physical underpinning, this `spot' model was partly
  motivated by analogy with magnetic oblique rotators; of course, a
  centred oblique dipole would be expected to give rise to {\em two}
  `spots', one of which would be visible at any phase.}
described by
\[
f_\alpha = f_0 + 
A(\cos{i}\cos{\beta} + \sin{i}\sin{\beta}\cos{\phi})(1 - u + u\cos\mu),
\]
where $f_0, A$ are normalizing constants, $i$ is the inclination of
the rotation axis to the line of sight, $\beta$ is the colatitude of
the spot, $\phi$ is the rotational phase, $u$ is a linear limb-darkening
coefficient, and $\mu$ is the angle between the line of sight and the
surface normal at the spot.  We find that the form of the \Ha\
light-curve can again be well matched by this model, 
with $u \simeq 0.6$, $i + \beta \simeq 105^\circ$, $i \simeq
\beta$ (Fig.~\ref{fig_toy}).  


\section{Summary and conclusions}

We have shown that the O6.5f?pe--O8fp spectroscopic variations
observed in HD~191612 are underpinned by an extremely regular
\mbox{538-d} `clock' which has kept good time across 24~years of
quantitative data (and which is documented a decade further back by
photographic material).  The Balmer and \hei\ lines show changes which
are highly reproducible on this period, and which are characterized by
variations in slightly redshifted, moderately narrow `excess'
emission.  Absorption lines of metals and \heii\ are essentially
constant in line strength (as are many selective emission lines), but
show radial-velocity changes arising in a double-lined
spectroscopic-binary orbit with $P_{\rm orb}= 1542$d.  The components'
properties are broadly in accord with an $\sim$O8 giant-like primary
and a $\sim$B1 main-sequence companion.

The results are entirely consistent with the \mbox{538-d} variability
arising through rotational modulation of a magnetically-constrained
plasma, as proposed by \citeauthor{Donati06a} (\citeyear{Donati06b}a).
As in the case of the B0.2$\;$V star $\tau$~Sco
(\citeauthor{Donati06b} \citeyear{Donati06b}b), the implied slow
rotation and the long-term stability of the variations argue that the
field originates in a fossil remnant, rather than a dynamo.  However,
although toy `disk' models (among others) are consistent with some
aspects of the data, we are unable to constrain the geometry of the
emission region in an interesting way. Future work could therefore
usefully concentrate on improving our understanding of the field
geometry, which should provide a better framework for numerical models
of optical emission, and perhaps help resolve the issues of
anomalously broad X-ray lines and soft X-ray emission reported by
\citet{Naze07}.  A continuing programme of circular spectropolarimetry
is working towards this aim, and a magnetic-field measurement from the
second-epoch ESPaDOnS observation used here already gives a clearly
detected field with a longitudinal component larger than the discovery
measurement, as expected for the geometry adopted by
\citeauthor{Donati06a} (\citeyear{Donati06b}a) and in our toy disk
model.

\section*{Acknowledgements}

We thank all the observers who contributed spectra to this campaign;
the service observers at the Isaac Newton Group; Sergio Ilovaisky and
Ashok K. Pati for orchestrating spectroscopy at OHP and VBO; Roger
Griffin for comments on the orbital solutions; and our referee, Otmar
Stahl, and his colleagues at the Landessternwarte Heidelberg for
establishing the date of Peppel's observation from the original logs.
We also gratefully acknowledge the following for support: the FNRS,
PRODEX/Belspo, and OPTICON (YN, GR); the Estonian Science Foundation,
Grant 6810 (KA, IK); and the F.H.~Levinson fund of the Peninsula
Community Foundation (DM).


\bibliography{IDH}

\begin{thebibliography}{}

\bibitem[\protect\citeauthoryear{{Babel} \& {Montmerle}}{{Babel} \&
  {Montmerle}}{1997}]{Babel97}
{Babel} J.,  {Montmerle} T.,  1997, \apjl, 485, L29

\bibitem[\protect\citeauthoryear{{Cardelli}, {Clayton} \& {Mathis}}{{Cardelli}
  et~al.}{1989}]{Cardelli89}
{Cardelli} J.~A.,  {Clayton} G.~C.,    {Mathis} J.~S.,  1989, \apj, 345, 245

\bibitem[\protect\citeauthoryear{{Donati}, {Babel}, {Harries}, {Howarth},
  {Petit} \& {Semel}}{{Donati} et~al.}{2002}]{Donati02}
{Donati} J.-F.,  {Babel} J.,  {Harries} T.~J.,  {Howarth} I.~D.,  {Petit} P.,
   {Semel} M.,  2002, \mnras, 333, 55

\bibitem[\protect\citeauthoryear{{Donati}, {Howarth}, {Bouret}, {Petit},
  {Catala} \& {Landstreet}}{{Donati} et~al.}{2006}]{Donati06a}
{Donati} J.-F.,  {Howarth} I.~D.,  {Bouret} J.-C.,  {Petit} P.,  {Catala} C.,
   {Landstreet} J.,  2006, \mnras, 365, L6

\bibitem[\protect\citeauthoryear{{Donati}, {Howarth}, {Jardine}, {Petit},
  {Catala}, {Landstreet}, {Bouret}, {Alecian}, {Barnes}, {Forveille}, {Paletou}
  \& {Manset}}{{Donati} et~al.}{2006}]{Donati06b}
{Donati} J.-F.,  {Howarth} I.~D.,  {Jardine} M.~M.,  {Petit} P.,  {Catala} C.,
  {Landstreet} J.~D.,  {Bouret} J.-C.,  {Alecian} E.,  {Barnes} J.~R.,
  {Forveille} T.,  {Paletou} F.,    {Manset} N.,  2006, \mnras, 370, 629

\bibitem[\protect\citeauthoryear{{Evans}, {Howarth}, {Irwin}, {Burnley} \&
  {Harries}}{{Evans} et~al.}{2004}]{Evans04}
{Evans} C.~J.,  {Howarth} I.~D.,  {Irwin} M.~J.,  {Burnley} A.~W.,    {Harries}
  T.~J.,  2004, \mnras, 353, 601

\bibitem[\protect\citeauthoryear{{Fullerton}, {Gies} \& {Bolton}}{{Fullerton}
  et~al.}{1996}]{Fullerton96}
{Fullerton} A.~W.,  {Gies} D.~R.,    {Bolton} C.~T.,  1996, \apjs, 103, 475

\bibitem[\protect\citeauthoryear{{Gagn{\'e}}, {Oksala}, {Cohen}, {Tonnesen},
  {ud-Doula}, {Owocki}, {Townsend} \& {MacFarlane}}{{Gagn{\'e}}
  et~al.}{2005}]{Gagne05}
{Gagn{\'e}} M.,  {Oksala} M.~E.,  {Cohen} D.~H.,  {Tonnesen} S.~K.,  {ud-Doula}
  A.,  {Owocki} S.~P.,  {Townsend} R.~H.~D.,    {MacFarlane} J.~J.,  2005,
  \apj, 628, 986

\bibitem[\protect\citeauthoryear{{Gray}}{{Gray}}{1992}]{Gray92}
{Gray} D.~F.,  1992, {The observation and analysis of stellar photospheres}.
Cambridge University Press

\bibitem[\protect\citeauthoryear{{Howarth}}{{Howarth}}{2003}]{Howarth03}
{Howarth} I.~D.,  2003, in {Maeder} A.,  {Eenens} P.,  eds, ASP Conf. Ser. 215:
  IAU Symp: Stellar Rotation p.~33

\bibitem[\protect\citeauthoryear{{Humphreys}}{{Humphreys}}{1978}]{Humphreys78}
{Humphreys} R.~M.,  1978, \apjs, 38, 309

\bibitem[\protect\citeauthoryear{{Kaper}, {Henrichs}, {Fullerton}, {Ando},
  {Bjorkman}, {Gies}, {Hirata}, {Kambe}, {McDavid} \& {Nichols}}{{Kaper}
  et~al.}{1997}]{Kaper97}
{Kaper} L.,  {Henrichs} H.~F.,  {Fullerton} A.~W.,  {Ando} H.,  {Bjorkman}
  K.~S.,  {Gies} D.~R.,  {Hirata} R.,  {Kambe} E.,  {McDavid} D.,    {Nichols}
  J.~S.,  1997, \aap, 327, 281

\bibitem[\protect\citeauthoryear{{Koen} \& {Eyer}}{{Koen} \&
  {Eyer}}{2002}]{Koen02}
{Koen} C.,  {Eyer} L.,  2002, \mnras, 331, 45

\bibitem[\protect\citeauthoryear{{Lanz} \& {Hubeny}}{{Lanz} \&
  {Hubeny}}{2003}]{Lanz03}
{Lanz} T.,  {Hubeny} I.,  2003, \apjs, 146, 417

\bibitem[\protect\citeauthoryear{{Massey} \& {Duffy}}{{Massey} \&
  {Duffy}}{2001}]{Massey01}
{Massey} P.,  {Duffy} A.~S.,  2001, \apj, 550, 713

\bibitem[\protect\citeauthoryear{{Moore}}{{Moore}}{1945}]{Moore45}
{Moore} C.~E.,  1945, A Multiplet Table of Astrophysical Interest, Princeton
  University Observatory, 20, D1

\bibitem[\protect\citeauthoryear{{Morel}, {Marchenko}, {Pati}, {Kuppuswamy},
  {Carini}, {Wood} \& {Zimmerman}}{{Morel} et~al.}{2004}]{Morel04}
{Morel} T.,  {Marchenko} S.~V.,  {Pati} A.~K.,  {Kuppuswamy} K.,  {Carini}
  M.~T.,  {Wood} E.,    {Zimmerman} R.,  2004, \mnras, 351, 552

\bibitem[\protect\citeauthoryear{{Naz\'{e}}}{{Naz\'{e}}}{2004}]{Naze04}
{Naz\'{e}} Y.,  2004, PhD thesis, Universit\'{e} de Li\`{e}ge

\bibitem[\protect\citeauthoryear{{Naz{\'e}}, {Rauw}, {Pollock}, {Walborn} \&
  {Howarth}}{{Naz{\'e}} et~al.}{2007}]{Naze07}
{Naz{\'e}} Y.,  {Rauw} G.,  {Pollock} A.~M.~T.,  {Walborn} N.~R.,    {Howarth}
  I.~D.,  2007, \mnras, 375, 145

\bibitem[\protect\citeauthoryear{{Peppel}}{{Peppel}}{1984}]{Peppel84}
{Peppel} U.,  1984, \aaps, 57, 107

\bibitem[\protect\citeauthoryear{{Puls}, {Urbaneja}, {Venero}, {Repolust},
  {Springmann}, {Jokuthy} \& {Mokiem}}{{Puls} et~al.}{2005}]{Puls05}
{Puls} J.,  {Urbaneja} M.~A.,  {Venero} R.,  {Repolust} T.,  {Springmann} U.,
  {Jokuthy} A.,    {Mokiem} M.~R.,  2005, \aap, 435, 669

\bibitem[\protect\citeauthoryear{{Repolust}, {Puls} \& {Herrero}}{{Repolust}
  et~al.}{2004}]{Repolust04}
{Repolust} T.,  {Puls} J.,    {Herrero} A.,  2004, \aap, 415, 349

\bibitem[\protect\citeauthoryear{{Stahl}, {Kaufer}, {Rivinius}, {Szeifert},
  {Wolf}, {Gaeng}, {Gummersbach}, {Jankovics}, {Kovacs}, {Mandel}, {Pakull} \&
  {Peitz}}{{Stahl} et~al.}{1996}]{Stahl96}
{Stahl} O.,  {Kaufer} A.,  {Rivinius} T.,  {Szeifert} T.,  {Wolf} B.,  {Gaeng}
  T.,  {Gummersbach} C.~A.,  {Jankovics} I.,  {Kovacs} J.,  {Mandel} H.,
  {Pakull} M.~W.,    {Peitz} J.,  1996, \aap, 312, 539

\bibitem[\protect\citeauthoryear{{ud-Doula} \& {Owocki}}{{ud-Doula} \&
  {Owocki}}{2002}]{udDoula02}
{ud-Doula} A.,  {Owocki} S.~P.,  2002, \apj, 576, 413

\bibitem[\protect\citeauthoryear{{ud-Doula}, {Townsend} \& {Owocki}}{{ud-Doula}
  et~al.}{2006}]{udDoula06}
{ud-Doula} A.,  {Townsend} R.~H.~D.,    {Owocki} S.~P.,  2006, \apjl, 640, L191

\bibitem[\protect\citeauthoryear{{Underhill}}{{Underhill}}{1995}]{Underhill95}
{Underhill} A.~B.,  1995, \apjs, 100, 433

\bibitem[\protect\citeauthoryear{{Wade}, {Fullerton}, {Donati}, {Landstreet},
  {Petit} \& {Strasser}}{{Wade} et~al.}{2006}]{Wade06}
{Wade} G.~A.,  {Fullerton} A.~W.,  {Donati} J.-F.,  {Landstreet} J.~D.,
  {Petit} P.,    {Strasser} S.,  2006, \aap, 451, 195

\bibitem[\protect\citeauthoryear{{Walborn}}{{Walborn}}{2001}]{Walborn01}
{Walborn} N.,  2001, in {Gull} T.~R.,  {Johannson} S.,   {Davidson} K.,  eds,
  ASP Conf. Ser. 242: Eta Carinae and Other Mysterious Stars: The Hidden
  Opportunities of Emission Spectroscopy p.~217

\bibitem[\protect\citeauthoryear{{Walborn}}{{Walborn}}{1973}]{Walborn73}
{Walborn} N.~R.,  1973, \aj, 78, 1067

\bibitem[\protect\citeauthoryear{{Walborn}}{{Walborn}}{2002}]{Walborn02}
{Walborn} N.~R.,  2002, \aj, 124, 507

\bibitem[\protect\citeauthoryear{{Walborn} \& {Fitzpatrick}}{{Walborn} \&
  {Fitzpatrick}}{2000}]{WalbornF00}
{Walborn} N.~R.,  {Fitzpatrick} E.~L.,  2000, \pasp, 112, 50

\bibitem[\protect\citeauthoryear{{Walborn}, {Howarth}, {Herrero} \&
  {Lennon}}{{Walborn} et~al.}{2003}]{Walborn03}
{Walborn} N.~R.,  {Howarth} I.~D.,  {Herrero} A.,    {Lennon} D.~J.,  2003,
  \apj, 588, 1025

\bibitem[\protect\citeauthoryear{{Walborn}, {Howarth}, {Rauw}, {Lennon},
  {Bond}, {Negueruela}, {Naz{\'e}}, {Corcoran}, {Herrero} \&
  {Pellerin}}{{Walborn} et~al.}{2004}]{Walborn04}
{Walborn} N.~R.,  {Howarth} I.~D.,  {Rauw} G.,  {Lennon} D.~J.,  {Bond} H.~E.,
  {Negueruela} I.,  {Naz{\'e}} Y.,  {Corcoran} M.~F.,  {Herrero} A.,
  {Pellerin} A.,  2004, \apjl, 617, L61

\bibitem[\protect\citeauthoryear{{Walborn}, {Lennon}, {Heap}, {Lindler},
  {Smith}, {Evans} \& {Parker}}{{Walborn} et~al.}{2000}]{Walborn00}
{Walborn} N.~R.,  {Lennon} D.~J.,  {Heap} S.~R.,  {Lindler} D.~J.,  {Smith}
  L.~J.,  {Evans} C.~J.,    {Parker} J.~W.,  2000, \pasp, 112, 1243

\bibitem[\protect\citeauthoryear{{Walborn} \& {Nichols}}{{Walborn} \&
  {Nichols}}{1994}]{Walborn94}
{Walborn} N.~R.,  {Nichols} J.~S.,  1994, \apjl, 425, L29

\end{thebibliography}

\bibliographystyle{mn2e}

\appendix
\section{Observational details}

This appendix summarizes
\begin{itemize}
\item
The log of observations (Table~\ref{tab_obs}, available in
full on-line)
\item
Radial-velocity measurements used in the orbital solutions
(Table~\ref{tab_rvs})
\end{itemize}

\begin{table*}
\caption[]{Format of Table~A1 (available in full on-line), the log of
  optical spectroscopy used in this paper.  Each `observation'
is a dataset collected at a single site in a given night, and may
consist of several separate spectra.  Phases are calculated with the
ephemerides of Table~\ref{tab_orbit} and eqtn.~\ref{eqn_ephem};  the notation
$\overline{C}.ppp$ means phase $+.ppp$ in cycle $-C$.  
Where two observers are identified,
the second name is the PI or instigator of the observation;  `srv'
indicates
a service-mode observation.   Where
observers provided smoothed or binned spectra, the effective
resolution is tabulated ($\Delta\lambda$).}
\begin{tabular}{lcrrr ll l l lr}
\hline
\multicolumn{1}{c}{JD}  &Year& 
\multicolumn{1}{c}{Phase $\phi_\alpha$} & 
\multicolumn{1}{c}{Phase $\varphi$} & 
\multicolumn{2}{c}{$\lambda$ range} &
Telescope/ &
Observer &
$\Delta\lambda$ &
\multicolumn{1}{c}{$W_\lambda$(\Ha)} 
\\
&&\multicolumn{1}{c}{(\Ha)}&\multicolumn{1}{c}{(orb)}
&\multicolumn{2}{c}{(nm)}&$\;$Instrument & &(\AA) &
\multicolumn{1}{c}{(\AA)} \\
\hline
   2447691.5 &  1989.45  &     $\overline{11}.35$ &       $\overline{4}.09$   &               & 651--662     & INT/IDS        &  Prinja/Howarth   &             0.5  &+1.37\\
   2447724.7 &  1989.54  &     $\overline{11}.42$ &       $\overline{4}.11$   &      404--499 &              & INT/IDS        &  Herrero          &             0.8  \\
   2447726.5 &  1989.55  &     $\overline{11}.42$ &       $\overline{4}.11$   &               & 634--677     & INT/IDS        &  Herrero          &             1.5  &+1.41\\
   2448117.4 &  1990.62  &     $\overline{10}.15$ &       $\overline{4}.37$   &      394--474 &              & INT/IDS        &  Howarth          &             1.5  \\ 
   2449139.7 &  1993.41  &      $\overline{8}.05$ &       $\overline{3}.03$   &      439--481 &              & INT/IDS        &  Vilchez          &             0.5  \\
\multicolumn{1}{c}{\vdots}  &
\vdots&
\vdots$\phantom{05}$ & 
\vdots$\phantom{15}$ & 
\multicolumn{1}{c}{\vdots}&
\multicolumn{1}{c}{\vdots}&
$\quad\;$\vdots &
$\quad$\vdots &
$\phantom{0}$\vdots&
\multicolumn{1}{c}{\vdots$\;\;$} 
\\
   2453955.4 &  2006.60 &       1.00 &                   0.15		   &               & 637--673     & Tartu          &  Kolka            & 1.5                  &$-4.00$\\
   2453956.5 &  2006.60 &       1.01 &                   0.15		   &               & 637--673     & Tartu          &  Kolka            & 1.5                  &$-3.96$\\
   2453958.4 &  2006.61 &       1.01 &                   0.15		   &               & 637--673     & Tartu          &  Kolka            & 1.5                  &$-4.14$\\
   2453966.6 &  2006.63 &       1.03 &                   0.16 		   &\multicolumn{2}{c}{384--706}  & WHT/ISIS       & Leisy/Lennon     & 0.4                  &$-4.07$\\
   2454002.5 &  2006.73 &       1.09 &                   0.18              &\multicolumn{2}{c}{384--706}  & WHT/ISIS       & Leisy/Lennon     & 0.4                  &$-3.32$\\
\hline
&&$\phantom{\overline{11}.365}$ &$\phantom{\overline{11}.365}$ &&&$\phantom{\mbox{CFHT/ESPADONS}}$&$\phantom{\mbox{RoeloefsX/Howarth}}$&$\phantom{.037,0.066      0.1}$\\
\end{tabular}
\label{tab_obs}
\end{table*}

\begin{table*}
\caption[]{Velocities used in orbital solutions.
Note that each 
of the three datasets is subject to a separate velocity zero-point
error of order a few~\kms\ (see footnote~\ref{foot_rvv}).
}
\begin{tabular}{lrc lrc lrc lrc lr}
\hline
\multicolumn{1}{c}{JD}  & \multicolumn{1}{c}{$V_\odot$}  & &
\multicolumn{1}{c}{JD}  & \multicolumn{1}{c}{$V_\odot$}  & &
\multicolumn{1}{c}{JD}  & \multicolumn{1}{c}{$V_\odot$}  & &
\multicolumn{1}{c}{JD}  & \multicolumn{1}{c}{$V_\odot$}  & &
\multicolumn{1}{c}{JD}  & \multicolumn{1}{c}{$V_\odot$}  \\
& \multicolumn{1}{c}{\kms}  &&
& \multicolumn{1}{c}{\kms}  &&
& \multicolumn{1}{c}{\kms}  &&
& \multicolumn{1}{c}{\kms}  &&
& \multicolumn{1}{c}{\kms}  \\
\hline
\hline
\multicolumn{4}{l}{$\qquad$\civ~5800{\AA}}\\
\hline
2449529.4 &  $-9.3$ &$\quad$& 2453203.5 & $-11.3$ &$\quad$& 2453275.2 & $-11.3$ &$\quad$& 2453474.6 &  $-8.8$ &$\quad$& 2453601.4 &  $+4.4$  \\
2452127.5 &  $+9.0$ &$\quad$& 2453225.8 &  $-6.3$ &$\quad$& 2453276.2 & $-10.6$ &$\quad$& 2453488.6 &  $-7.3$ &$\quad$& 2453625.4 &  $+3.7$  \\
2452528.8 &  $-8.3$ &$\quad$& 2453246.3 &  $-8.9$ &$\quad$& 2453278.3 & $-11.7$ &$\quad$& 2453519.6 &  $-5.9$ &$\quad$& 2453652.3 &  $+7.0$  \\
2452549.4 &  $-3.1$ &$\quad$& 2453251.3 &  $-9.6$ &$\quad$& 2453282.4 & $-12.0$ &$\quad$& 2453536.7 &  $-5.8$ &$\quad$& 2453653.3 &  $+7.6$  \\
2452607.3 &  $-5.3$ &$\quad$& 2453264.3 &  $-7.8$ &$\quad$& 2453283.3 & $-10.1$ &$\quad$& 2453545.0 &  $-3.2$ &$\quad$& 2453668.3 &  $+6.3$  \\
2452801.6 & $-11.4$ &$\quad$& 2453265.3 &  $-9.6$ &$\quad$& 2453302.3 & $-11.3$ &$\quad$& 2453546.0 &  $-3.7$ &$\quad$& 2453681.3 & $+11.7$  \\
2452808.5 & $-11.2$ &$\quad$& 2453266.3 & $-12.2$ &$\quad$& 2453314.6 & $-10.3$ &$\quad$& 2453547.0 &  $-3.5$ &$\quad$& 2453836.8 &  $+6.3$  \\
2452816.6 & $-11.0$ &$\quad$& 2453267.2 &  $-9.9$ &$\quad$& 2453316.3 &  $-9.7$ &$\quad$& 2453548.0 &  $-4.3$ &$\quad$& 2453892.7 &  $-0.2$  \\
2452871.2 & $-12.1$ &$\quad$& 2453267.4 &  $-8.2$ &$\quad$& 2453320.3 & $-12.9$ &$\quad$& 2453554.6 &  $-7.4$ &$\quad$& 2453896.0 &  $+2.1$  \\
2453132.9 &  $-6.5$ &$\quad$& 2453269.2 & $-10.9$ &$\quad$& 2453324.3 & $-12.7$ &$\quad$& 2453568.5 &  $-0.8$ &$\quad$& 2453941.5 &  $-3.4$  \\
2453146.7 & $-10.4$ &$\quad$& 2453270.2 & $-14.2$ &$\quad$& 2453326.3 & $-13.2$ &$\quad$& 2453585.5 &  $-2.2$ &$\quad$& 2453966.6 &  $-1.7$  \\
2453192.5 &  $-7.6$ &$\quad$& 2453272.4 &  $-8.3$ &$\quad$& 2453346.4 &  $-9.5$ &$\quad$& 2453594.5 &  $-4.5$ &$\quad$& 2454002.5 &  $-3.2$  \\   
\hline
\hline
\multicolumn{4}{l}{$\qquad\lambda$6700 emission}\\
\hline
2450683.5 &  $-9.4$ &$\quad$& 2452808.5 & $-34.8$ &$\quad$& 2453266.6 & $-36.6$ &$\quad$& 2453474.6 & $-33.3$ &$\quad$& 2453594.5 & $-22.0$   \\
2452127.4 & $-16.4$ &$\quad$& 2452919.3 & $-34.6$ &$\quad$& 2453270.2 & $-30.2$ &$\quad$& 2453480.7 & $-29.6$ &$\quad$& 2453607.4 & $-15.6$   \\
2452128.4 & $-16.2$ &$\quad$& 2453132.9 & $-42.5$ &$\quad$& 2453272.4 & $-31.4$ &$\quad$& 2453488.6 & $-33.6$ &$\quad$& 2453625.4 & $-24.2$   \\
2452129.4 & $-17.6$ &$\quad$& 2453146.7 & $-38.8$ &$\quad$& 2453278.3 & $-37.8$ &$\quad$& 2453519.6 & $-26.6$ &$\quad$& 2453652.3 & $-18.2$   \\
2452129.6 & $-15.3$ &$\quad$& 2453157.7 & $-35.3$ &$\quad$& 2453283.3 & $-36.0$ &$\quad$& 2453536.7 & $-16.6$ &$\quad$& 2453653.3 & $-23.1$   \\
2452130.4 & $-15.6$ &$\quad$& 2453209.7 & $-37.2$ &$\quad$& 2453290.6 & $-34.7$ &$\quad$& 2453545.0 & $-28.0$ &$\quad$& 2453668.3 & $-16.5$   \\
2452132.7 & $-18.3$ &$\quad$& 2453210.5 & $-37.6$ &$\quad$& 2453291.6 & $-36.2$ &$\quad$& 2453546.0 & $-28.0$ &$\quad$& 2453681.3 & $-23.1$   \\
2452483.6 & $-29.9$ &$\quad$& 2453225.8 & $-33.2$ &$\quad$& 2453326.3 & $-41.4$ &$\quad$& 2453547.0 & $-27.8$ &$\quad$& 2453836.8 & $-16.1$   \\
2452566.3 & $-29.2$ &$\quad$& 2453246.3 & $-40.0$ &$\quad$& 2453368.3 & $-31.4$ &$\quad$& 2453548.0 & $-27.2$ &$\quad$& 2453892.7 & $-27.2$   \\
2452607.3 & $-32.0$ &$\quad$& 2453250.5 & $-38.2$ &$\quad$& 2453369.3 & $-37.7$ &$\quad$& 2453554.6 & $-23.3$ &$\quad$& 2453896.0 & $-17.5$   \\
2452801.6 & $-32.6$ &$\quad$& 2453251.3 & $-41.5$ &$\quad$& 2453446.7 & $-39.0$ &$\quad$& 2453585.5 & $-27.5$ &$\quad$& 2453966.6 & $-25.8$   \\
\hline
\hline
\multicolumn{4}{l}{$\qquad$\oii\ absorption}\\
\hline
2452101.6 & $-28.7$ &$\quad$& 2452566.3 & $-24.6$ &$\quad$& 2453157.7 & $+21.6$ &$\quad$& 2453266.6 & $+19.7$ &$\quad$& 2453681.3 & $-35.1$   \\
2452127.4 & $-33.0$ &$\quad$& 2452632.3 &  $-5.9$ &$\quad$& 2453209.7 & $+15.7$ &$\quad$& 2453554.6 &  $-6.6$ &$\quad$& 2453836.8 & $-32.7$   \\
2452127.5 & $-33.5$ &$\quad$& 2452808.5 &  $+7.2$ &$\quad$& 2453210.5 &  $+6.9$ &$\quad$& 2453568.5 & $-11.1$ &$\quad$& 2453892.7 & $-21.4$   \\
2452128.4 & $-33.4$ &$\quad$& 2452816.6 &  $+5.7$ &$\quad$& 2453225.8 & $+11.3$ &$\quad$& 2453625.4 & $-26.3$ &$\quad$& 2453896.0 & $-23.4$   \\
2452129.4 & $-32.7$ &$\quad$& 2452831.6 &  $+6.1$ &$\quad$& 2453246.3 &  $+5.0$ &$\quad$& 2453652.3 & $-19.1$ &$\quad$& 2453941.5 & $-17.2$   \\
2452130.4 & $-34.9$ &$\quad$& 2452837.6 & $+14.3$ &$\quad$& 2453250.5 & $+11.6$ &$\quad$& 2453653.3 & $-25.8$ &$\quad$& 2453966.6 & $-20.4$   \\
\hline
\end{tabular}
\label{tab_rvs}
\end{table*}

\newpage
\addtocounter{table}{-2}
\begin{table*}
\caption[]{Optical spectroscopy used in this paper.  Each `observation'
is a dataset collected at a single site in a given night, and may
consist of several separate spectra.  Phases are calculated with the
ephemerides of Table~2 and eqtn.~2;  the notation
$\overline{C}.ppp$ means phase $+.ppp$ in cycle $-C$.  
Where two observers are identified,
the second name is the PI or instigator of the observation;  `srv'
indicates
a service-mode observation.   Where
observers provided smoothed or binned spectra, the effective
resolution is tabulated ($\Delta\lambda$).}
\begin{tabular}{lcrrr ll l l lr}
\hline
\multicolumn{1}{c}{JD}  &Year& 
\multicolumn{1}{c}{Phase $\phi_\alpha$} & 
\multicolumn{1}{c}{Phase $\varphi$} & 
\multicolumn{2}{c}{$\lambda$ range} &
Telescope/ &
Observer &
$\Delta\lambda$ &
\multicolumn{1}{c}{$W_\lambda$(\Ha)} 
\\
&&\multicolumn{1}{c}{(\Ha)}&\multicolumn{1}{c}{(orb)}
&\multicolumn{2}{c}{(nm)}&$\;$Instrument & &(\AA) &
\multicolumn{1}{c}{(\AA)} \\
\hline
   2447691.5 &  1989.45  &     $\overline{11}.35$ &       $\overline{4}.09$   &               & 651--662     & INT/IDS        &  Prinja/Howarth   &             0.5  &+1.37\\
   2447724.7 &  1989.54  &     $\overline{11}.42$ &       $\overline{4}.11$   &      404--499 &              & INT/IDS        &  Herrero          &             0.8  \\
   2447726.5 &  1989.55  &     $\overline{11}.42$ &       $\overline{4}.11$   &               & 634--677     & INT/IDS        &  Herrero          &             1.5  &+1.41\\
   2448117.4 &  1990.62  &     $\overline{10}.15$ &       $\overline{4}.37$   &      394--474 &              & INT/IDS        &  Howarth          &             1.5  \\ 
   2449139.7 &  1993.41  &      $\overline{8}.05$ &       $\overline{3}.03$   &      439--481 &              & INT/IDS        &  Vilchez          &             0.5  \\
   2449529.4 &  1994.48  &      $\overline{8}.77$ &       $\overline{3}.28$   &      445--603 &              & WHT/ISIS       &  Crowther         &             1.5  \\
   2450683.5 &  1997.64  &      $\overline{6}.92$ &       $\overline{2}.03$   &      356--407 & 641--677     & WHT/UES        &  Trapero          &0.07, 0.13            &$-3.82$\\
   2452101.6 &  2001.52  &      $\overline{3}.56$ &       $\overline{2}.95$   &      369--535 &              & INT/IDS        &  Erwin/Herrero    & 0.20                 \\
   2452127.4 &  2001.59  &      $\overline{3}.60$ &       $\overline{2}.97$   &      388--482 & 636--676     & WHT/ISIS       &  Evans/Howarth    & 0.5, 0.8            &+1.58\\ 
   2452127.5 &  2001.59  &      $\overline{3}.60$ &       $\overline{2}.97$   &      361--599 &              & WHT/UES        &  Evans/Howarth    &0.06 		   \\
   2452128.4 &  2001.60  &      $\overline{3}.61$ &       $\overline{2}.97$   &      388--477 & 636--676     & WHT/ISIS       &  Evans/Howarth    & 0.5, 0.8 	   &+1.59\\
   2452129.4 &  2001.60  &      $\overline{3}.61$ &       $\overline{2}.97$   &      388--478 & 636--676     & WHT/ISIS       &  Evans/Howarth    & 0.5, 0.8 	   &+1.51\\
   2452129.6 &  2001.60  &      $\overline{3}.61$ &       $\overline{2}.97$   &      365--543 & 636--676     & WHT/ISIS       &  Evans/Howarth    & 0.8, 0.8     &+1.46\\
   2452130.4 &  2001.60  &      $\overline{3}.61$ &       $\overline{2}.97$   &      388--478 & 636--676     & WHT/ISIS       &  Evans/Howarth    & 0.5, 0.8 	   &+1.49\\
   2452132.7 &  2001.61  &      $\overline{3}.61$ &       $\overline{2}.97$   &               & 635--677     & INT/IDS        &  Srv/Herrero   & 0.6             	   &+1.35\\
   2452483.6 &  2002.57  &      $\overline{2}.27$ &       $\overline{1}.20$   &               & 625--681     & INT/IDS        &  Srv/Herrero   & 0.6                  &+0.39\\
   2452508.2 &  2002.64  &      $\overline{2}.31$ &       $\overline{1}.21$   &\multicolumn{2}{c}{322--703}  & INT/IDS        &  Araujo/Gansicke  & 3.8                  &+0.79\\
   2452528.8 &  2002.69  &      $\overline{2}.35$ &       $\overline{1}.23$   &\multicolumn{2}{c}{320--733}  & Bok            &  Wagner/Bond      & 1.6,            2.6  &+1.19\\
   2452549.4 &  2002.75  &      $\overline{2}.39$ &       $\overline{1}.24$   &\multicolumn{2}{c}{440--775}  & WHT/ISIS       &  Rix/Pettini      & 1.8--2.6		   &+1.34\\
   2452566.3 &  2002.80  &      $\overline{2}.42$ &       $\overline{1}.25$   &      395--485 & 633--708     & WHT/ISIS       &  Crowther         & 0.8       	   &+1.42\\
   2452607.3 &  2002.91  &      $\overline{2}.50$ &       $\overline{1}.28$   &\multicolumn{2}{c}{382--797}  & WHT/ISIS       &  Evans/Crowther   & 0.8,           3.0  &+1.33\\           
   2452632.3 &  2002.98  &      $\overline{2}.54$ &       $\overline{1}.29$   &      405--495 &              & WHT/ISIS       &  Harries          & 0.7                  \\
   2452775.0 &  2003.37  &      $\overline{2}.81$ &       $\overline{1}.39$   &      405--473 &              & WIYN           &  Bond             & 0.6                  \\
   2452801.6 &  2003.44  &      $\overline{2}.86$ &       $\overline{1}.40$   &\multicolumn{2}{c}{386--684}  & WHT/ISIS       &  Lennon/Herrero   & 0.6--1.2             &$-2.70$\\
   2452803.7 &  2003.45  &      $\overline{2}.86$ &       $\overline{1}.41$   &      425--472 &              & INT/IDS        &  Christian        & 0.3                  \\
   2452808.5 &  2003.46  &      $\overline{2}.87$ &       $\overline{1}.41$   &\multicolumn{2}{c}{380--700}  & WHT/ISIS       &  Herrero          & 0.7 		   &$-2.66$\\
   2452810.8 &  2003.47  &      $\overline{2}.88$ &       $\overline{1}.41$   &      405--473 &              & WIYN           &  Bond             & 3.9     		   \\
   2452816.6 &  2003.48  &      $\overline{2}.89$ &       $\overline{1}.41$   &\multicolumn{2}{c}{330--684}  & INT/IDS        &  Lennon           & 0.6       		   &$-2.67$\\
   2452831.6 &  2003.52  &      $\overline{2}.91$ &       $\overline{1}.42$   &      372--519 & 599-697      & INT/IDS        &  Prada            & 0.7 		   &$-3.00$\\
   2452832.5 &  2003.53  &      $\overline{2}.92$ &       $\overline{1}.42$   &      332--498 & 586--705     & INT/IDS        &  Howarth          & 1.4, 1.1      &$-3.69$\\
   2452833.4 &  2003.53  &      $\overline{2}.92$ &       $\overline{1}.42$   &               & 587--704     & INT/IDS        &  Howarth          & 1.1             	   &$-3.92$\\
   2452834.7 &  2003.53  &      $\overline{2}.92$ &       $\overline{1}.43$   &      353--521 &              & INT/IDS        &  Howarth          & 1.1 		   \\
   2452835.7 &  2003.53  &      $\overline{2}.92$ &       $\overline{1}.43$   &      353--522 & 585--697     & INT/IDS        &  Howarth          & 1.4, 1.1             &$-3.79$\\
   2452836.7 &  2003.54  &      $\overline{2}.92$ &       $\overline{1}.43$   &      352--522 &              & INT/IDS        &  Howarth          & 1.4 		   \\
   2452837.6 &  2003.54  &      $\overline{2}.93$ &       $\overline{1}.43$   &      382--490 & 630--683     & INT/IDS        &  Howarth          & 1.1, 0.5  	   &$-3.74$\\
   2452838.6 &  2003.54  &      $\overline{2}.93$ &       $\overline{1}.43$   &      382--491 & 627--683     & INT/IDS        &  Howarth          & 1.1, 0.5  	   &$-3.15$\\
   2452845.9 &  2003.56  &      $\overline{2}.94$ &       $\overline{1}.43$   &      405--473 &              & WIYN           &  Harmer/Bond      & 0.6 		   \\
   2452854.8 &  2003.59  &      $\overline{2}.96$ &       $\overline{1}.44$   &      405--473 &              & WIYN           &  Harmer/Bond      & 0.6 		   \\
   2452871.2 &  2003.63  &      $\overline{2}.99$ &       $\overline{1}.45$   &\multicolumn{2}{c}{391--716}  & OMM            &  Pellerin         & 3.9		   &$-4.30$\\
   2452902.6 &  2003.72  &      $\overline{1}.05$ &       $\overline{1}.47$   &      405--473 &              & WIYN           &  Bond             & 0.6 		   \\
   2452913.7 &  2003.75  &      $\overline{1}.07$ &       $\overline{1}.48$   &      334--648 &              & MMT            &  Bond             & 3.6                  \\
   2452919.3 &  2003.76  &      $\overline{1}.08$ &       $\overline{1}.48$   &               & 634--677     & OHP/Aurelie    &  Rauw             & 0.6                  &$-3.61$\\
   2452922.4 &  2003.77  &      $\overline{1}.08$ &       $\overline{1}.48$   &      446--490 &              & OHP/Aurelie    &      Rauw         & 0.6 		   \\
   2452980.3 &  2003.93  &      $\overline{1}.19$ &       $\overline{1}.52$   &      373--547 & 576--756     & WHT/ISIS       &  Lennon/Herrero   & 1.2 		   &$-1.40$\\
   2453132.9 &  2004.35  &      $\overline{1}.47$ &       $\overline{1}.62$   &\multicolumn{2}{c}{444--674}  & OHP/Elodie     &  Rauw             & 0.3                 &+1.52\\
   2453132.9 &  2004.35  &      $\overline{1}.47$ &       $\overline{1}.62$   &      405--474 &              & WIYN           &  Harmer/Bond      & 0.6     		   \\
   2453146.7 &  2004.39  &      $\overline{1}.50$ &       $\overline{1}.63$   &      374--550 & 568--742     & WHT/ISIS       &  Licandro         & 1.6 		   &+1.30\\
   2453147.4 &  2004.39  &      $\overline{1}.50$ &       $\overline{1}.63$   &\multicolumn{2}{c}{371--693}  & Skinakas       &  Reig             & 2.0		   &+1.27\\
   2453157.7 &  2004.42  &      $\overline{1}.52$ &       $\overline{1}.64$   &      359--494 & 604--681     & WHT/ISIS       &  Howarth          & 0.8                  &+1.30\\
   2453180.4 &  2004.48  &      $\overline{1}.56$ &       $\overline{1}.65$   &               & 550--686     & Skinakas       &  Reig             & 2.0		   &+1.30\\
   2453182.4 &  2004.48  &      $\overline{1}.57$ &       $\overline{1}.65$   &      375--510 &              & Skinakas       &  Reig             & 2.0		   \\
   2453192.5 &  2004.51  &      $\overline{1}.59$ &       $\overline{1}.66$   &\multicolumn{2}{c}{390--682}  & OHP/Elodie     &  Siviero          & 0.1                  &+1.29\\
   2453193.4 &  2004.51  &      $\overline{1}.59$ &       $\overline{1}.66$   &               & 551--687     & Skinakas       &  Reig             & 2.0		   &+1.45\\
   2453194.4 &  2004.52  &      $\overline{1}.59$ &       $\overline{1}.66$   &      377--509 &              & Skinakas       &  Reig             & 2.0		   \\
   2453202.5 &  2004.54  &      $\overline{1}.60$ &       $\overline{1}.66$   &      350--530 & 609--818     & Loiano         &  Negueruela       & 3.0, 3.0             &+1.50\\
   2453203.5 &  2004.54  &      $\overline{1}.61$ &       $\overline{1}.66$   &\multicolumn{2}{c}{388--999}  & Loiano         &  Negueruela       & 0.7 		   &+1.47\\
   2453209.7 &  2004.56  &      $\overline{1}.62$ &       $\overline{1}.67$   &      449--540 & 597--688     & WHT/ISIS       &  R.C. Smith       & 0.7                  &+1.36\\
   2453210.5 &  2004.56  &      $\overline{1}.62$ &       $\overline{1}.67$   &      449--540 & 597--689     & WHT/ISIS       &  R.C. Smith       & 0.7 		   &+1.34\\
&&$\phantom{\overline{11}.365}$ &$\phantom{\overline{11}.365}$ &&&$\phantom{\mbox{CFHT/ESPADONS}}$&$\phantom{\mbox{RoeloefsX/Howarth}}$&$\phantom{.037,0.066 0.1}$\\
\end{tabular}
\end{table*}

\begin{table*}
\begin{tabular}{lcrrr ll l l lr}
\hline
\multicolumn{1}{c}{JD}  & Year&
\multicolumn{1}{c}{Phase $\phi_\alpha$} & 
\multicolumn{1}{c}{Phase $\varphi$} & 
\multicolumn{2}{c}{$\lambda$ range} &
Telescope/ &
Observer &
$\Delta\lambda$ & \multicolumn{1}{c}{$W_\lambda$(\Ha)} 
\\
&&\multicolumn{1}{c}{(\Ha)}&\multicolumn{1}{c}{(orb)}
&\multicolumn{2}{c}{(nm)}&$\;$Instrument & &(\AA) &
\multicolumn{1}{c}{(\AA)}\\
\hline
   2453225.8 &  2004.60  &     $\overline{1}.65$ &       $\overline{1}.68$ &\multicolumn{2}{c}{372--684}  & SPM            &  Georgiev         & 0.4 		   &+1.31\\
   2453243.4 &  2004.65  &     $\overline{1}.68$ &       $\overline{1}.69$ &\multicolumn{2}{c}{362--684}  & Skinakas       &  Reig             & 2.0		   &+1.06\\
   2453246.3 &  2004.66  &     $\overline{1}.69$ &       $\overline{1}.69$ &      401--590 & 653--674     & OHP/Elodie     &  Srv/Rauw      & 0.3                 &+1.02\\
   2453250.5 &  2004.67  &     $\overline{1}.69$ &       $\overline{1}.70$ &      353--505 & 602--688     & WHT/ISIS       &  Ostensen/Howarth & 0.6                  &+0.89\\
   2453251.3 &  2004.67  &     $\overline{1}.70$ &       $\overline{1}.70$ &      401--590 & 653--674     & OHP/Elodie     &  Srv/Rauw         & 0.3                 &+0.89\\
   2453252.4 &  2004.67  &     $\overline{1}.70$ &       $\overline{1}.70$ &\multicolumn{2}{c}{365--683}  & Skinakas       &  Reig             & 2.0		   &+0.90\\
   2453261.3 &  2004.70  &     $\overline{1}.71$ &       $\overline{1}.70$ &      363--570 &              & Skinakas       &  Reig             & 2.0                  \\
   2453262.4 &  2004.70  &     $\overline{1}.72$ &       $\overline{1}.70$ &\multicolumn{2}{c}{363--684}  & Skinakas       &  Reig             & 2.0		   &+0.73\\
   2453264.3 &  2004.71  &     $\overline{1}.72$ &       $\overline{1}.70$ &\multicolumn{2}{c}{420--710}  & Crimea         &  Antokhin         & 0.5 		   &+0.72\\
   2453265.3 &  2004.71  &     $\overline{1}.72$ &       $\overline{1}.71$ &\multicolumn{2}{c}{420--710}  & Crimea         &  Antokhin         & 0.5 		   &+0.62\\
   2453266.3 &  2004.71  &     $\overline{1}.72$ &       $\overline{1}.71$ &\multicolumn{2}{c}{420--710}  & Crimea         &  Antokhin         & 0.5 		   &+0.53\\
   2453266.6 &  2004.71  &     $\overline{1}.72$ &       $\overline{1}.71$ &      346--505 & 597--689     & WHT/ISIS       &  Benn/Howarth     & 0.6 		   &+0.58\\
   2453267.2 &  2004.72  &     $\overline{1}.72$ &       $\overline{1}.71$ &\multicolumn{2}{c}{420--710}  & Crimea         &  Antokhin         & 0.5 		   &+0.58\\
   2453267.4 &  2004.72  &     $\overline{1}.73$ &       $\overline{1}.71$ &      401--590 & 653--674     & OHP/Elodie     &  Srv/Rauw    & 0.3                 &+0.60\\
   2453269.2 &  2004.72  &     $\overline{1}.73$ &       $\overline{1}.71$ &\multicolumn{2}{c}{420--710}  & Crimea         &  Antokhin         & 0.5                  &+0.62\\
   2453270.2 &  2004.72  &     $\overline{1}.73$ &       $\overline{1}.71$ &\multicolumn{2}{c}{420--710}  & Crimea         &  Antokhin         & 0.5 		   &+0.40\\
   2453272.4 &  2004.73  &     $\overline{1}.73$ &       $\overline{1}.71$ &      401--590 & 653--674     & OHP/Elodie     &  Srv/Rauw     & 0.3                 &+0.54\\
   2453275.2 &  2004.74  &     $\overline{1}.74$ &       $\overline{1}.71$ &\multicolumn{2}{c}{420--710}  & Crimea         &  Antokhin         & 0.5 		   &+0.30\\
   2453276.2 &  2004.74  &     $\overline{1}.74$ &       $\overline{1}.71$ &\multicolumn{2}{c}{420--710}  & Crimea         &  Antokhin         & 0.5 		   &+0.18\\
   2453278.3 &  2004.75  &     $\overline{1}.75$ &       $\overline{1}.71$ &      401--590 & 653--674     & OHP/Elodie     &  Srv/Rauw        & 0.3                 &+0.25\\
   2453282.4 &  2004.76  &     $\overline{1}.75$ &       $\overline{1}.72$ &\multicolumn{2}{c}{365--685}  & Skinakas       &  Reig             & 2.0		   \\
   2453282.4 &  2004.76  &     $\overline{1}.75$ &       $\overline{1}.72$ &\multicolumn{2}{c}{370--790}  & TNG/SARG       &  Andreuzzi        & 0.04		   &+0.22\\
   2453283.3 &  2004.76  &     $\overline{1}.75$ &       $\overline{1}.72$ &      401--590 & 653--674     & OHP/Elodie     &  Srv/Rauw    & 0.3                 &+0.18\\
   2453286.4 &  2004.77  &     $\overline{1}.76$ &       $\overline{1}.72$ &      445--490 &              & OHP/Aurelie    &      Rauw         & 0.6		   \\
   2453289.3 &  2004.78  &     $\overline{1}.77$ &       $\overline{1}.72$ &               & 636--675     & OHP/Aurelie    &      Rauw         & 0.6		   &$-0.27$\\
   2453290.6 &  2004.78  &     $\overline{1}.77$ &       $\overline{1}.72$ &               & 647--714     & KPNO           &  Gies             & 0.7 		   &$-0.42$\\
   2453291.6 &  2004.78  &     $\overline{1}.77$ &       $\overline{1}.72$ &               & 647--714     & KPNO           &  Gies             & 0.7 		   &$-0.31$\\
   2453293.3 &  2004.79  &     $\overline{1}.77$ &       $\overline{1}.72$ &      445--490 &              & OHP/Aurelie    &      Rauw         & 0.6 		   \\
   2453302.3 &  2004.81  &     $\overline{1}.79$ &       $\overline{1}.73$ &      401--590 & 653--674     & OHP/Elodie     &  Srv/Rauw     & 0.3                 &$-0.64$\\
   2453304.3 &  2004.82  &     $\overline{1}.79$ &       $\overline{1}.73$ &               & 644--662     & Skinakas       &  Reig             & 2.0                  &$-0.80$\\
   2453314.6 &  2004.84  &     $\overline{1}.81$ &       $\overline{1}.74$ &\multicolumn{2}{c}{366--668}  & SPM            &  Georgiev         & 0.3                  &$-1.44$\\
   2453316.3 &  2004.85  &     $\overline{1}.82$ &       $\overline{1}.74$ &      401--590 & 653--674     & OHP/Elodie     &  Srv/Rauw     & 0.3                 &$-1.44$\\
   2453320.3 &  2004.86  &     $\overline{1}.82$ &       $\overline{1}.74$ &      401--590 & 653--674     & OHP/Elodie     &  Srv/Rauw      & 0.3                 &$-1.60$\\
   2453324.3 &  2004.87  &     $\overline{1}.83$ &       $\overline{1}.74$ &\multicolumn{2}{c}{390--682}  & OHP/Elodie     &  Srv/Rauw      & 0.1                  &$-1.77$\\
   2453326.3 &  2004.88  &     $\overline{1}.83$ &       $\overline{1}.74$ &      401--590 & 653--674     & OHP/Elodie     &  Srv/Rauw      & 0.1 		   &$-1.85$\\
   2453346.4 &  2004.93  &     $\overline{1}.87$ &       $\overline{1}.76$ &\multicolumn{2}{c}{381--703}  & PicMidi/Musicos&  Negueruela       & 0.1                  &$-3.25$\\
   2453368.3 &  2004.99  &     $\overline{1}.91$ &       $\overline{1}.77$ &      360--537 & 613--704     & WHT/ISIS       &  Roeloefs/Howarth & 1.1, 0.6             &$-3.65$\\
   2453369.3 &  2004.99  &     $\overline{1}.91$ &       $\overline{1}.77$ &      360--537 & 613--704     & WHT/ISIS       &  Roeloefs/Howarth & 1.1, 0.6             &$-3.53$\\
   2453371.3 &  2005.00  &     $\overline{1}.92$ &       $\overline{1}.77$ &      353--502 & 611--777     & WHT/ISIS       &  Gaensicke        & 1.7,  3.2            &$-3.48$\\
   2453444.8 &  2005.20  &      0.06             &       $\overline{1}.82$ &\multicolumn{2}{c}{364--834}  & NOT/ALFOSC     &  Licandro         & 2.8		   &$-3.90$\\
   2453446.7 &  2005.21  &      0.06             &       $\overline{1}.82$ &               & 639--705     & Asiago         &  Morel            & 1.2                  &$-3.53$\\
   2453458.5 &  2005.24  &      0.08             &       $\overline{1}.83$ &               & 637--673     & Tartu          &  Annuk/Kolka      & 1.5                  &$-3.70$\\
   2453465.6 &  2005.26  &      0.09             &       $\overline{1}.83$ &      401--590 & 653--674     & OHP/Elodie     &  Srv/Rauw       & 0.3                 &$-3.22$\\
   2453472.8 &  2005.28  &      0.11             &       $\overline{1}.84$ &               & 550--700     & FMO            &  McDavid          & 1.8 		   &$-3.29$\\
   2453474.6 &  2005.28  &      0.11             &       $\overline{1}.84$ &      401--590 & 653--674     & OHP/Elodie     &  Srv/Rauw    & 0.3                 &$-3.01$\\
   2453480.7 &  2005.30  &      0.12             &       $\overline{1}.84$ &      350--505 & 600--689     & WHT/ISIS       &  Licandro         & 0.6                  &$-2.96$\\   
   2453486.9 &  2005.32  &      0.13             &       $\overline{1}.85$ &               & 560--680     & FMO            &  McDavid          & 1.8                  &$-2.80$\\
   2453488.6 &  2005.32  &      0.14             &       $\overline{1}.85$ &      401--590 & 653--674     & OHP/Elodie     &  Srv/Rauw        & 0.3                 &$-2.42$\\
   2453515.4 &  2005.39  &      0.19             &       $\overline{1}.87$ &               & 509--716     & Skinakas       &  Reig             & 2.0                  &$-1.49$\\
   2453519.6 &  2005.41  &      0.19             &       $\overline{1}.87$ &      401--590 & 653--674     & OHP/Elodie     &  Srv/Rauw       & 0.3                 &$-1.08$\\
   2453536.7 &  2005.45  &      0.23             &       $\overline{1}.88$ &      401--590 & 653--674     & OHP/Elodie     &  Srv/Rauw        & 0.3                 &$-0.44$\\
   2453545.0 &  2005.48  &      0.24             &       $\overline{1}.89$ &\multicolumn{2}{c}{$\phantom{0}$369--1048}  & CFHT/ESPaDOnS  &  Donati           & 0.06		   \\
   2453545.4 &  2005.48  &      0.24             &       $\overline{1}.89$ &               & 510--715     & Skinakas       &  Reig             & 2.0		   \\
   2453546.0 &  2005.48  &      0.24             &       $\overline{1}.89$ &\multicolumn{2}{c}{$\phantom{0}$369--1048}  & CFHT/ESPaDOnS  &  Donati           & 0.06		  \\
   2453547.0 &  2005.48  &      0.25             &       $\overline{1}.89$ &\multicolumn{2}{c}{$\phantom{0}$369--1048}  & CFHT/ESPaDOnS  &  Donati           & 0.06                 &$-0.14$\\
   2453548.0 &  2005.48  &      0.25             &       $\overline{1}.89$ &\multicolumn{2}{c}{$\phantom{0}$369--1048}  & CFHT/ESPaDOnS  &  Donati           & 0.06		   \\
   2453554.6 &  2005.50  &      0.26             &       $\overline{1}.89$ &      401--590 & 653--674     & OHP/Elodie     &  Srv/Rauw        & 0.3                 &+0.37\\
   2453559.5 &  2005.52  &      0.27             &       $\overline{1}.90$ &\multicolumn{2}{c}{391--999}  & Loiano         &  Negueruela       & 1.5                  &+0.48\\
   2453563.4 &  2005.53  &      0.28             &       $\overline{1}.90$ &\multicolumn{2}{c}{468--716}  & Skinakas       &  Reig             & 2.0		   &+0.51\\
   2453565.5 &  2005.53  &      0.28             &       $\overline{1}.90$ &               & 510--716     & Skinakas       &  Reig             & 2.0		   &+0.63\\
   2453568.5 &  2005.54  &      0.29             &       $\overline{1}.90$ &      401--590 & 653--674     & OHP/Elodie     &  Srv/Rauw     & 0.3                 &+0.66\\
   2453585.5 &  2005.59  &      0.32             &       $\overline{1}.91$ &\multicolumn{2}{c}{390--682}  & OHP/Elodie     &  Srv/Rauw        & 0.1                  &+0.94\\
   2453594.5 &  2005.61  &      0.33             &       $\overline{1}.92$ &      401--590 & 653--674     & OHP/Elodie     &  Srv/Rauw        & 0.3                 &+1.22\\
&&$\phantom{\overline{11}.365}$ &$\phantom{\overline{11}.365}$ &&&$\phantom{\mbox{CFHT/ESPADONS}}$&$\phantom{\mbox{RoeloefsX/Howarth}}$&$\phantom{.037,0.066      0.1}$\\

\end{tabular}
\end{table*}

\begin{table*}
\begin{tabular}{lcrrr ll l l lr}
\hline
\multicolumn{1}{c}{JD}  & Year &
\multicolumn{1}{c}{Phase $\phi_\alpha$} & 
\multicolumn{1}{c}{Phase $\varphi$} & 
\multicolumn{2}{c}{$\lambda$ range} &
Telescope/ &
Observer &
$\Delta\lambda$ & \multicolumn{1}{c}{$W_\lambda$(\Ha)} \\
&&\multicolumn{1}{c}{(\Ha)}&\multicolumn{1}{c}{(orb)}
&\multicolumn{2}{c}{(nm)}&$\;$Instrument & &(\AA) &
\multicolumn{1}{c}{(\AA)}\\
\hline
   2453600.3 &  2005.63 &       0.34 &                   $\overline{1}.92$ &\multicolumn{2}{c}{362--715}  & Skinakas       &  Reig             & 2.0                  &+1.29\\
   2453601.5 &  2005.63 &       0.35 &                   $\overline{1}.92$ &\multicolumn{2}{c}{389--682}  & OHP/Elodie     &  Bouret/Negueruela& 0.3                 &+1.39\\
   2453603.9 &  2005.64 &       0.35 &                   $\overline{1}.92$ &               & 646--662     & DAO            &  Bohlender        & 0.3                  &+1.50\\
   2453606.9 &  2005.64 &       0.36 &                   $\overline{1}.93$ &               & 647--662     & DAO            &  Bohlender        & 0.3                  &+1.37\\
   2453607.4 &  2005.65 &       0.36 &                   $\overline{1}.93$ &               & 648--684     & Tartu          &  Kolka            & 1.5                  &+1.38\\
   2453609.8 &  2005.65 &       0.36 &                   $\overline{1}.93$ &               & 647--662     & DAO            &  Bohlender        & 0.3                  &+1.30\\
   2453615.4 &  2005.67 &       0.37 &                   $\overline{1}.93$ &               & 637--673     & Tartu          &  Kolka            & 1.5                  &+1.38\\
   2453616.4 &  2005.67 &       0.37 &                   $\overline{1}.93$ &               & 649--662     & Tartu          &  Kolka            & 0.5                  &+1.54\\
   2453620.4 &  2005.68 &       0.38 &                   $\overline{1}.94$ &               & 649--662     & Tartu          &  Kolka            & 0.5                  &+1.45\\
   2453621.4 &  2005.68 &       0.38 &                   $\overline{1}.94$ &               & 637--673     & Tartu          &  Kolka            & 1.5                  &+1.42\\
   2453625.4 &  2005.70 &       0.39 &                   $\overline{1}.94$ &      401--590 & 653--674     & OHP/Elodie     &  Srv/Rauw        & 0.3                 &+1.52\\
   2453627.4 &  2005.70 &       0.39 &                   $\overline{1}.94$ &               & 637--674     & Tartu          &  Kolka            & 1.5                  &+1.49\\
   2453630.4 &  2005.71 &       0.40 &                   $\overline{1}.94$ &               & 649--662     & Tartu          &  Kolka            & 0.5                  &+1.38\\
   2453631.4 &  2005.71 &       0.40 &                   $\overline{1}.94$ &               & 637--674     & Tartu          &  Kolka            & 1.5                  &+1.46\\
   2453652.3 &  2005.77 &       0.44 &                   $\overline{1}.96$ &      401--590 & 653--674     & OHP/Elodie     &  Srv/Rauw      & 0.3                 &+1.48\\
   2453653.3 &  2005.77 &       0.44 &                   $\overline{1}.96$ &      401--590 & 653--674     & OHP/Elodie     &  Srv/Rauw      & 0.3                 &+1.55\\
   2453653.3 &  2005.77 &       0.44 &                   $\overline{1}.96$ &               & 649--662     & Tartu          &  Kolka            & 0.5                  &+1.36\\
   2453668.3 &  2005.81 &       0.47 &                   $\overline{1}.97$ &\multicolumn{2}{c}{390--682}  & OHP/Elodie     &  Srv/Rauw        & 0.03                 &+1.35\\
   2453681.3 &  2005.85 &       0.49 &                   $\overline{1}.97$ &      401--590 & 653--674     & OHP/Elodie     &  Srv/Rauw      & 0.3                 &+1.58\\
   2453718.6 &  2005.95 &       0.56 &                   0.00		   &               & 647--662     & DAO            &  Bohlender        & 0.3                  &+1.37\\
   2453722.6 &  2005.96 &       0.57 &                   0.00		   &               & 647--662     & DAO            &  Bohlender        & 0.3                  &+1.42\\
   2453777.1 &  2006.11 &       0.67 &                   0.04		   &               & 648--673     & DAO            &  Bohlender        & 0.3                  &+1.03\\
   2453782.1 &  2006.12 &       0.68 &                   0.04		   &               & 648--662     & DAO            &  Bohlender        & 0.3                  &+0.96\\
   2453795.7 &  2006.16 &       0.71 &                   0.05		   &               & 649--684     & Tartu          &  Kolka            & 1.5                  &+1.00\\
   2453810.6 &  2006.20 &       0.74 &                   0.06		   &               & 649--684     & Tartu          &  Kolka            & 1.5                  &+0.44\\
   2453817.6 &  2006.22 &       0.75 &                   0.06		   &               & 637--673     & Tartu          &  Kolka            & 1.5                  &+0.14\\
   2453836.8 &  2006.27 &       0.78 &                   0.08		   &\multicolumn{2}{c}{385--707}  & WHT/ISIS       &  Lennon           & 0.6                  &$-0.68$\\
   2453845.5 &  2006.30 &       0.80 &                   0.08		   &               & 637--673     & Tartu          &  Kolka            & 1.5                  &$-1.18$\\
   2453860.4 &  2006.34 &       0.83 &                   0.09		   &               & 637--673     & Tartu          &  Kolka            & 1.5                  &$-1.87$\\
   2453863.4 &  2006.35 &       0.83 &                   0.09		   &               & 637--673     & Tartu          &  Kolka            & 1.5                  &$-2.08$\\
   2453865.4 &  2006.35 &       0.84 &                   0.09		   &               & 637--673     & Tartu          &  Kolka            & 1.5                  &$-2.12$\\
   2453867.4 &  2006.36 &       0.84 &                   0.10		   &               & 637--673     & Tartu          &  Kolka            & 1.5                  &$-2.09$\\
   2453871.5 &  2006.37 &       0.85 &                   0.10		   &               & 637--673     & Tartu          &  Kolka            & 1.5                  &$-2.32$\\
   2453879.4 &  2006.39 &       0.86 &                   0.10		   &               & 637--673     & Tartu          &  Kolka            & 1.5                  &$-2.47$\\
   2453889.5 &  2006.42 &       0.88 &                   0.11		   &               & 637--673     & Tartu          &  Kolka            & 1.5                  &$-3.00$\\
   2453892.7 &  2006.43 &       0.89 &                   0.11		   &\multicolumn{2}{c}{386--705}  & WHT/ISIS       &  Lennon           & 0.4                  &$-3.11$\\
   2453895.4 &  2006.44 &       0.89 &                   0.11		   &               & 637--673     & Tartu          &  Kolka            & 1.5                  &$-3.19$\\
   2453896.0 &  2006.44 &       0.89 &                   0.11		   &\multicolumn{2}{c}{$\phantom{0}$369--1048} & CFHT/ESPaDOnS & Donati  & 0.06                 &$-3.17$\\
   2453900.4 &  2006.45 &       0.90 &                   0.12		   &               & 637--673     & Tartu          &  Kolka            & 1.5                  &$-3.15$\\
   2453941.5 &  2006.56 &       0.98 &                   0.14		   &\multicolumn{2}{c}{384--706}  & WHT/ISIS       & Leisy/Lennon     & 0.4                  &$-4.10$\\
   2453955.4 &  2006.60 &       1.00 &                   0.15		   &               & 637--673     & Tartu          &  Kolka            & 1.5                  &$-4.00$\\
   2453956.5 &  2006.60 &       1.01 &                   0.15		   &               & 637--673     & Tartu          &  Kolka            & 1.5                  &$-3.96$\\
   2453958.4 &  2006.61 &       1.01 &                   0.15		   &               & 637--673     & Tartu          &  Kolka            & 1.5                  &$-4.14$\\
   2453966.6 &  2006.63 &       1.03 &                   0.16 		   &\multicolumn{2}{c}{384--706}  & WHT/ISIS       & Leisy/Lennon     & 0.4                  &$-4.07$\\
   2454002.5 &  2006.73 &       1.09 &                   0.18              &\multicolumn{2}{c}{384--706}  & WHT/ISIS       & Leisy/Lennon     & 0.4                  &$-3.32$\\
\hline
\hline
&&$\phantom{\overline{11}.365}$ &$\phantom{\overline{11}.365}$ &&&$\phantom{\mbox{CFHT/ESPADONS}}$&$\phantom{\mbox{RoeloefsX/Howarth}}$&$\phantom{.037,0.066      0.1}$\\
\end{tabular}
\end{table*}

\end{document}